\documentclass[10pt,aps,prd,reprint,floatfix,amsfonts,amssymb,amsmath,preprintnumbers,notitlepage,nofootinbib]{revtex4-1}
\usepackage[ascii]{inputenc}
\usepackage[british]{babel}
\usepackage[Symbolsmallscale]{upgreek}
\usepackage{isomath}
\usepackage[protrusion=true,expansion=true,tracking=true,kerning=true,spacing=true,final,babel=true]{microtype}
\usepackage{physics}
\usepackage{empheq}
\usepackage{xcolor}
\usepackage{url}
\usepackage{booktabs}
\usepackage[final]{hyperref}

\begin{document}
\title{Anisotropies in the astrophysical gravitational-wave background:\\Predictions for the detection of compact binaries by LIGO and Virgo}

\author{Alexander~C.~Jenkins}
\email{Alexander.Jenkins@kcl.ac.uk}
\affiliation{Theoretical Particle Physics and Cosmology Group, Physics Department, King's College London, University of London, Strand, London WC2R 2LS, United Kingdom}

\author{Mairi~Sakellariadou}
\email{Mairi.Sakellariadou@kcl.ac.uk}
\affiliation{Theoretical Particle Physics and Cosmology Group, Physics Department, King's College London, University of London, Strand, London WC2R 2LS, United Kingdom}

\author{Tania~Regimbau}
\email{Tania.Regimbau@oca.eu}
\affiliation{Universit\'e Savoie Mont Blanc, CNRS/IN2P3, Laboratoire d'Annecy-le-Vieux de Physique des Particules (LAPP), 74941 Annecy, France}
\affiliation{Universit\'e C\^ote d'Azur, Observatoire de la C\^ote d'Azur, CNRS, Laboratoire Artemis, CS 34229, 06304 Nice Cedex 4, France}

\author{Eric~Slezak}
\email{Eric.Slezak@oca.eu}
\affiliation{Universit\'e C\^ote d'Azur, Observatoire de la C\^ote d'Azur, CNRS, Laboratoire Lagrange, CS 34229, 06304 Nice Cedex 4, France}

\date{\today}
\preprint{KCL-PH-TH/2018-23}

\begin{abstract}
  We develop a detailed anisotropic model for the astrophysical gravitational-wave background, including binary mergers of two stellar-mass black holes, two neutron stars, or one of each, which are expected to be the strongest contributions in the LIGO-Virgo frequency band.
  The angular spectrum of the anisotropies, quantified by the $C_\ell$ components, is calculated using two complementary approaches: (i) a simple, closed-form analytical expression, and (ii) a detailed numerical study using an all-sky mock light cone galaxy catalogue from the Millennium simulation.
  The two approaches are in excellent agreement at large angular scales, and differ by a factor of order unity at smaller scales.
  These anisotropies are considerably larger in amplitude than e.g. those in the temperature of the cosmic microwave background, confirming that it is important to model these anisotropies, and indicating that this is a promising avenue for future theoretical and observational work.
\end{abstract}
\maketitle

\section{Introduction}
The first direct detections of gravitational waves (GW) by the Advanced LIGO~\cite{Harry:2010zz,TheLIGOScientific:2014jea} and Advanced Virgo~\cite{TheVirgo:2014hva} detectors, from the inspiral and merger of pairs of black holes~\cite{Abbott:2016nmj,Abbott:2017vtc,Abbott:2017gyy,Abbott:2017oio}, and the recent first detection of GWs from the inspiral and merger of a pair of neutron stars~\cite{TheLIGOScientific:2017qsa}, have opened a new window to the Universe, its physical processes, and the astrophysical sources that populate it.
Each of these detections are associated with individual loud events, but it is expected that there are many more quiet compact binary mergers in the Universe that are too faint or too distant to be individually resolved.
Signals from these quiet events superimpose to create a stochastic gravitational-wave background (SGWB), which is discernible from instrumental noise by cross-correlating the output from multiple detectors, and which is expected to be detected by Advanced LIGO and Advanced Virgo after a few years of operation at design sensitivity~\cite{Abbott:2016nmj}.
Many other sources, both astrophysical (e.g.,~rotating neutron stars, supernovae, core collapse formation of black holes or neutron stars) and cosmological (e.g.,~phase transitions, inflation, cosmic strings), are expected to contribute to the SGWB.
The cosmological background can provide unique information about the early stages of our Universe, while the astrophysical background probes the Universe's evolution since the beginning of star formation.

Gravitational-wave sources with an inhomogeneous spatial distribution lead to a SGWB characterized by preferred directions, and therefore anisotropies.
This is analogous to the measured temperature anisotropies in the cosmic microwave background (CMB) radiation.
Methods to detect these SGWB anisotropies using radiometer and spherical harmonic techniques have been developed~\cite{Ballmer:2005uw,Abbott:2007tw,Thrane:2009fp,Abadie:2011rr,Messenger:2015tga,Chung:2011da,Sun:2016idj,TheLIGOScientific:2013cya} and applied to data from Advanced LIGO's first observational run to set upper limits on the energy flux and energy density of the SGWB as a function of sky position~\cite{TheLIGOScientific:2016xzw}.
Recently, Refs.~\cite{Cusin:2017fwz,Cusin:2017mjm,Jenkins:2018nty} have investigated a general framework to model the anisotropies in the SGWB induced by various types of sources, which was applied to the case of cosmic strings in Ref.~\cite{Jenkins:2018nty} and to the case of binary black hole (BBH) mergers in Ref.~\cite{Cusin:2018rsq}.

In what follows we use a result [Eq.~(26)] of Ref.~\cite{Jenkins:2018nty} to study anisotropies induced in the SGWB by astrophysical sources, namely from compact object mergers within anisotropically distributed galaxies.
We assume that the background is stationary, which implies that the statistical properties of the SGWB do not vary over the time scale of the observation.
Our analysis follows two distinct approaches.
In the former (analytical approach), we use simple analytical functions for the galaxy number density and the galaxy-galaxy two-point correlation function to derive a simple, closed-form expression for the two-point correlation function and its multipole components $C_\ell$.
Let us point out that while Ref.~\cite{Cusin:2018rsq} also used a two-point correlation function, our study differs in several ways: (i) we include all three different types of merger: binaries consisting of two neutron stars, two stellar-mass black holes, or one of each; (ii) we calculate the kinematic dipole, which we then subtract since it interferes with the anisotropy statistics (as demonstrated in Ref.~\cite{Jenkins:2018nty}); (iii) we follow the fiducial model of the LIGO/Virgo collaboration paper~\cite{Abbott:2017xzg} rather than the astrophysical model of Refs.~\cite{Dvorkin:2017kfg,Dvorkin:2016okx,Dvorkin:2016wac} to estimate the rate of mergers; (iv) we use a nonlinear power law expression Eq.~\eqref{eq:galaxy-2pcf} to model the two-point galaxy clustering, rather than basing this on linear transfer functions of the matter overdensity as in Ref.~\cite{Cusin:2018rsq}.
In the latter (catalogue approach), we make use---for the first time in the SGWB literature---of a realistic mock catalogue of galaxies and employ recipes to infer the production and merger rates of compact objects.

In both approaches, we restrict our attention to the most important sources for LIGO, Virgo, and other planned ground-based detectors.
Future space-based detectors such as the Laser Interferometer Space Antenna (LISA)~\cite{Audley:2017drz} will probe a lower-frequency window of the GW spectrum, and will therefore be sensitive to different populations of astrophysical systems, which will produce a different stochastic background signal.
However, much of Sec.~\ref{sec:AGWB} is still valid in this case, and we intend to repeat our analysis for the LISA frequency band in a future work.

Our present study is organized as follows.
In Sec.~\ref{sec:AGWB} we derive a general expression for the astrophysical gravitational wave background (AGWB) induced by an anisotropic distribution of galaxies hosting compact binary mergers.
We first write the SGWB density parameter $\Omega_\mathrm{gw}$ in terms of the average number density of galaxies per comoving volume, the galaxy number overdensity, the rate of binary mergers per galaxy, the observer's peculiar velocity, and the GW strain spectrum of each binary merger.
These quantities depend on the star formation rates (SFRs) and metallicities of the galaxies, and the spins and masses of the two binary components.
We perform our analysis using the fiducial model of Ref.~\cite{Abbott:2017xzg}, and thus consider the merger rate, the distribution of binary parameters, and the gravitational waveforms emitted by the binary mergers as those of Ref.~\cite{Abbott:2017xzg}.
We then decompose the density parameter as the average (isotropic) part $\bar{\Omega}_\mathrm{gw}$ and the GW density contrast $\delta_\mathrm{gw}$, which encodes the anisotropies.
In Sec.~\ref{sec:analytical} we use analytical functions to describe the average galaxy number density and the galaxy-galaxy two-point correlation function and derive a simple closed-form expression for each of the $C_\ell$ coefficients, up to a frequency-dependent factor which we evaluate numerically.
In Sec.~\ref{sec:catalogue} we follow a more accurate approach and use a realistic mock catalogue of galaxies, hence relaxing some of the assumptions made in the analytical approach.
In Sec.~\ref{sec:results-discussion} we compare the analytical and numerical approaches, and comment on the imprint to the SGWB from the anisotropic distribution of compact object mergers as compared to those inferred by a cosmic string network~\cite{Jenkins:2018nty}.
Some technical details are given explicitly in the appendixes.

\section{General expression for the AGWB}
\label{sec:AGWB}
The dimensionless density parameter expressing the intensity of a SGWB, with observed frequency between $\nu_\mathrm{o}$ and $\nu_\mathrm{o}+\dd{\nu_\mathrm{o}}$ and arriving from an infinitesimal solid angle $\dd[2]{\sigma_\mathrm{o}}$ centered on the direction $\vu*e_\mathrm{o}$, is defined as
    \begin{equation}
        \Omega_\mathrm{gw}\qty(\nu_\mathrm{o},\vu*e_\mathrm{o})\equiv\frac{1}{\rho_\mathrm{c}}\frac{\dd[3]{\rho_\mathrm{gw}}}{\dd(\ln\nu_\mathrm{o})\dd[2]{\sigma_\mathrm{o}}}
        =\frac{8\uppi G\nu_\mathrm{o}}{3H_\mathrm{o}^2}\frac{\dd[3]{\rho_\mathrm{gw}}}{\dd{\nu_\mathrm{o}}\dd[2]{\sigma_\mathrm{o}}},
    \end{equation}
    where we have used the customary normalization with respect to the critical density $\rho_\mathrm{c}=3H_0^2/(8\uppi G)$.
To study anisotropies in the SGWB induced by astrophysical sources, we consider a Friedmann-Lema\^{i}tre-Robertson-Walker spacetime, and neglect cosmological perturbations, keeping only the anisotropy due to the source density contrast and the dipole induced by the peculiar motion of the observer.
Following Ref.~\cite{Jenkins:2018nty}, we thus have
    \begin{align}
    \begin{split}
        \label{eq:omega}
        \Omega_\mathrm{gw}\qty(\nu_\mathrm{o},\vu*e_\mathrm{o})=&\frac{\uppi}{3}\qty(t_H\nu_\mathrm{o})^3\int_0^\infty\dd{z}\frac{1+z}{E\qty(z)}\\
        &\times\int\dd{\vb*\zeta}\bar{n}R\qty(1+\delta_n+\vu*e_\mathrm{o}\vdot\vb*v_\mathrm{o})\int_{S^2}\dd[2]{\sigma_\mathrm{s}}r\mathrlap{^2}_\mathrm{s}\,\tilde{h}^2,
    \end{split}
    \end{align}
    where $t_H\equiv1/H_\mathrm{o}$ is the Hubble time, $z$ is the redshift, $\vb*\zeta$ represents the set of source parameters, $\bar{n}$ is the average (homogeneous) source number density per comoving unit volume, $R$ is the rate of gravitational wave bursts per source, $\delta_n\equiv(n-\bar{n})/\bar{n}$ is the source number overdensity, $\vu*e_\mathrm{o}$ is the observation direction, $\vb*v_\mathrm{o}$ is the observer's peculiar velocity, $\tilde{h}$ is the GW strain spectrum of a burst, and the final integral is over a sphere centered on the source.
Note that Eq.~(\ref{eq:omega}) has been modified with respect to Eq.~(26) of Ref.~\cite{Jenkins:2018nty} by integrating over redshift rather than conformal time, and by using the number density per comoving volume, rather than per physical volume.
Assuming the standard flat $\Lambda$CDM cosmology, we have
    \begin{equation}
        E\qty(z)\equiv\frac{H\qty(z)}{H_\mathrm{o}}=\sqrt{\Omega_\mathrm{m}\qty(1+z)^3+\Omega_\Lambda},
    \end{equation}
    with $\Omega_\mathrm{m}=0.3065$, $\Omega_\Lambda=0.6935$, and $H_\mathrm{o}=67.9\text{ km s}^{-1}\text{Mpc}^{-1}$.

Note that, while Eq.~(26) of Ref.~\cite{Jenkins:2018nty} imposed a cutoff time on the line-of-sight integral to remove nearby, loud, infrequent, resolvable sources and isolate the stochastic part of the signal (see Sec.~IIC of Ref.~\cite{Jenkins:2018nty}), this is not done in Eq.~\eqref{eq:omega} above.
This is because the background from binary mergers consists of far fewer GW-emitting events than the cosmic string background considered in Ref.~\cite{Jenkins:2018nty}.
As a result, the problem of distinguishing different events from each other is greatly reduced, and the detector noise becomes the main barrier to resolving events individually.
Indeed, a network of future ``third-generation" detectors is expected to be able to resolve the vast majority of binary mergers in this frequency window, including more than 99.9\% of BBH mergers~\cite{Regimbau:2016ike,Smith:2017vfk,Abbott:2017xzg}.
(For a discussion of how such a large ensemble of individually resolved mergers may be used as a cosmological probe, see Refs.~\cite{Namikawa:2015prh,Namikawa:2016edr}.)
The astrophysical background from unresolvable binary mergers therefore depends on the sensitivity of the detector network, and future detectors will be able to greatly reduce the level of the background by resolving binaries out to much larger redshifts.
In order to phrase our results in a detector-independent way, we have opted to include \emph{all} binary merger events, no matter how close they are to the observer.

In what follows, we study the imprint of an anisotropic distribution of binary mergers, within galaxies, on the SGWB.
Since any anisotropies on scales smaller than the typical size of a galaxy are inaccessible to us, we treat galaxies as point sources.
We therefore interpret $n$ as the number density of galaxies, and $R$ as the rate of binary mergers per galaxy.
Let us write
    \begin{equation}
        \vb*\zeta=\qty(\vb*\zeta_\mathrm{b},\vb*\zeta_\mathrm{g}),
    \end{equation}
    where $\vb*\zeta_\mathrm{g}$ are the parameters of the galaxy (mass, age, luminosity, metallicity, etc.) and $\vb*\zeta_\mathrm{b}$ are the parameters of the compact binary (masses and spins of the components).
We then have
    \begin{equation}
        n=n\qty(z,\vu*e_\mathrm{o},\vb*\zeta_\mathrm{g}),\quad\tilde{h}=\tilde{h}\qty(\nu_\mathrm{s},\vu*e_\mathrm{s},\vb*\zeta_\mathrm{b}),\quad R=R\qty(z,\vb*\zeta_\mathrm{g},\vb*\zeta_\mathrm{b}),
    \end{equation}
    since the galaxy number density is independent of the source parameters, the strain is independent of the galaxy parameters, and the rate per galaxy depends on both.
Note that $\tilde{h}$ is a function of the source-frame frequency $\nu_\mathrm{s}$, which is related to the observed frequency $\nu_\mathrm{o}$ by
    \begin{equation}
        \label{eq:source-frequency}
        \nu_\mathrm{s}=\nu_\mathrm{o}\qty(1+z)\qty[1+\vu*e_\mathrm{o}\vdot\qty(\vb*v_\mathrm{g}-\vb*v_\mathrm{o})],
    \end{equation}
    where $\vb*v_\mathrm{g}$ and $\vb*v_\mathrm{o}$ are the peculiar velocities of the galaxy and the observer, respectively.
We allow each source to emit GWs anisotropically, so that $\tilde{h}$ is also a function of the direction of propagation away from the source's position, $\vu*e_\mathrm{s}$.

We can capture much of the relevant astrophysical information with just six parameters: the SFR $\psi$ and metallicity $Z$ of the galaxies, and the masses and spins of the two binary components.
The compact objects we consider are all the end products of stellar evolution, so we can use the SFR history of a galaxy to calculate its population of compact binaries, and therefore parameterize the rate of mergers of these objects.
The formation of massive black holes from stars is inhibited by stellar winds in high-metallicity environments, so we also include the galaxy metallicity as a parameter to account for this.
All other parameters describing the galactic environment are neglected.
We further assume that the compact binary orbits are circular and have nonprecessing spins (this is expected to be the case for mergers resulting from isolated binary evolution---such binaries are expected to circularize long before merger due to gravitational backreaction).

Let $\psi\qty(z)$ be the SFR of a galaxy at redshift $z$ in units of $M_\odot\text{ yr}^{-1}$, and $Z$ be the metallicity of the galaxy (i.e. the fraction of the galaxy's mass that is in elements heavier than helium).
In order to use the SFR to parameterize the merger rate, we must take into account the time delay between stars being formed, evolving to become compact objects, and eventually merging with each other.
These delay times are typically much longer than the timescales over which the SFR of a galaxy changes, so cannot be neglected.
We therefore define the \emph{delayed} SFR,
    \begin{equation}
        \psi_\mathrm{d}\qty(z)\equiv\int_0^\infty\dd{t_\mathrm{d}}p\qty(t_\mathrm{d})\psi\qty(z_\mathrm{f}),
    \end{equation}
    where $z_\mathrm{f}\qty(z,t_\mathrm{d})$ is the redshift at which the stars are formed, at a time $t_\mathrm{d}$ before the merger occurs at redshift $z$.
This is the convolution of the SFR with the probability distribution for the delay times, which varies depending on the objects we consider.
Certainly, not all stars become merging compact objects, so $\psi_\mathrm{d}$ is not equal to the merger rate---but it \emph{is} proportional to it.
It is therefore much simpler to use $\psi_\mathrm{d}$ as a galaxy parameter, rather than $\psi$.
Similarly, rather than using the metallicity $Z$, it is more convenient to use a logarithmic scaling, and to normalize relative to the solar metallicity $Z_\odot\approx0.02$. So we define
    \begin{equation}
        \mathcal{Z}\equiv\log_{10}\frac{Z}{Z_\odot},\qquad\mathcal{Z}\in(-\infty,\mathcal{Z}_\mathrm{max}],
    \end{equation}
    where $\mathcal{Z}_\mathrm{max}\equiv-\log_{10}Z_\odot\approx1.70$.

Hence we write the galaxy and binary parameter vectors as
    \begin{equation}
        \vb*\zeta_\mathrm{g}=\qty(\psi_\mathrm{d},\mathcal{Z}),\qquad\vb*\zeta_\mathrm{b}=\qty(m_1,m_2,\chi_1,\chi_2),
    \end{equation}
    where $\psi_\mathrm{d}$ and $\mathcal{Z}$ are the delayed SFR and log-normalized metallicity introduced above, $m_1$ and $m_2$ are the masses of the two compact objects in a binary (measured in units of $M_\odot$), and
    \begin{equation}
        \chi_1\equiv\frac{S_1}{m_1^2},\qquad\chi_2\equiv\frac{S_2}{m_2^2},\qquad\chi_1,\chi_2\in[-1,+1],
    \end{equation}
    are the dimensionless spin parameters of the compact objects (where $S_i$ is the spin angular momentum).
Using these parameters, we consider three different types of merger: binaries consisting of two neutron stars, two stellar-mass black holes, or one of each.
We call these BNS, BBH, and BHNS respectively.
Each type of binary has a different average merger rate, a different distribution over the parameters $\vb*\zeta_\mathrm{b}$, and a different distribution for the delay time $t_\mathrm{d}$ (giving a different delayed SFR $\psi_\mathrm{d}$).

Equation (\ref{eq:omega}) therefore becomes
    \begin{align}
    \begin{split}
        \label{eq:omega-gw}
        \Omega_\mathrm{gw}\qty(\nu_\mathrm{o},\vu*e_\mathrm{o})=&\sum_i\frac{\uppi}{3}\qty(t_H\nu_\mathrm{o})^3\int_0^{z_\mathrm{max}}\dd{z}\frac{1+z}{E\qty(z)}\\
        &\times\int\dd{\vb*\zeta_\mathrm{g}}\bar{n}\qty(1+\delta_n+\vu*e_\mathrm{o}\vdot\vb*v_\mathrm{o})\int\dd{\vb*\zeta_\mathrm{b}}R_i\mathcal{S}_i,
    \end{split}
    \end{align}
    where index $i$ runs over BBH, BNS, and BHNS, and for brevity we have introduced the variable
    \begin{equation}
        \mathcal{S}_i\qty(\nu_\mathrm{s},\vb*\zeta_\mathrm{b})\equiv\int_{S^2}\dd[2]{\sigma_\mathrm{s}}r\mathrlap{^2}_\mathrm{s}\,\tilde{h}_i^2.
    \end{equation}
We have set an upper limit $z_\mathrm{max}=10$ on the redshift integral, reflecting the fact that the merger rate is essentially 0 at redshifts greater than this.
The galaxy and binary parameter integration measures are
    \begin{equation}
        \int\dd{\vb*\zeta_\mathrm{g}}=\int_0^{+\infty}\dd{\psi_{\mathrm{d},i}}\int_{-\infty}^{\mathcal{Z}_\mathrm{max}}\dd{\mathcal{Z}},
    \end{equation}
    \begin{equation}
        \int\dd{\vb*\zeta_\mathrm{b}}=\int_0^\infty\dd{m_1}\int_0^\infty\dd{m_2}\int_{-1}^{+1}\dd{\chi_1}\int_{-1}^{+1}\dd{\chi_2},
    \end{equation}
    where we have allowed the delayed SFR to be different for different types of binary, by allowing a different distribution of delay times for each case,
    \begin{equation}
        \psi_{\mathrm{d},i}\qty(z)\equiv\int_0^\infty\dd{t_\mathrm{d}}p_i\qty(t_\mathrm{d})\psi(z_\mathrm{f}).
    \end{equation}
For the $\Lambda$CDM cosmology we consider, the formation redshift $z_\mathrm{f}$ of an object that takes a time $t_\mathrm{d}$ to eventually merge at redshift $z$ is given by
    \begin{align}
    \begin{split}
        \label{eq:z_f}
        &1+z_\mathrm{f}\qty(z,t_\mathrm{d})\\
        &=\qty(1+z)\qty[\cosh\qty(\frac{3\Omega_\Lambda^{1/2}t_\mathrm{d}}{2t_H})-\frac{E\qty(z)}{\Omega_\Lambda^{1/2}}\sinh\qty(\frac{3\Omega_\Lambda^{1/2}t_\mathrm{d}}{2t_H})]^{-2/3}.
    \end{split}
    \end{align}

\subsection{Intragalactic details of the model}
In what follows we describe  in detail how we model intragalactic quantities (i.e., those that are relevant within each galaxy)---the merger rate, the distribution of binary parameters, and the gravitational waveforms emitted by the binary mergers.
This is almost identical to the fiducial model in Ref.~\cite{Abbott:2017xzg}, to which we refer the reader for a more thorough discussion and justification of the various modeling choices.
The only differences from Ref.~\cite{Abbott:2017xzg} are (i) the inclusion of BHNS mergers, for completeness, and (ii) the use of the merger rate per galaxy, rather than per comoving volume (this simplifies our later analysis regarding the anisotropies in the background).
The modeling of intergalactic quantities---the galaxy number density $n$ and its clustering statistics---is different in our two approaches, and is discussed in Secs.~\ref{sec:analytical} and~\ref{sec:catalogue}.

First, in order to calculate $\mathcal{S}_i$, we use the hybrid waveform models of Refs.~\cite{Ajith:2007kx,Ajith:2009bn}.
These are valid for BBH, and when integrated over the sphere give
    \begin{align}
    \begin{split}
        \label{eq:waveforms}
        \mathcal{S}_\mathrm{BBH}&\equiv\int_{S^2}\dd[2]{\sigma_\mathrm{s}}r\mathrlap{^2}_\mathrm{s}\tilde{h}\mathrlap{^2}_\mathrm{BBH}\\
        &=\frac{5\qty(G\mathcal{M})^{5/3}}{6\uppi^{1/3}}\\
        &\times
        \begin{cases}
            \nu_\mathrm{s}^{-7/3}\qty[1+\sum_{i=2}^3\alpha_i\qty(\uppi GM\nu_\mathrm{s})^{i/3}]^2, & \nu_\mathrm{s}<\nu_1\\
            c_1\nu_\mathrm{s}^{-4/3}\qty[1+\sum_{i=1}^2\epsilon_i\qty(\uppi GM\nu_\mathrm{s})^{i/3}]^2, & \nu_1\le\nu_\mathrm{s}<\nu_2\\
            c_2\qty[1+\qty(\frac{\nu_\mathrm{s}-\nu_2}{\nu_3})^2]^{-2}, & \nu_2\le\nu_\mathrm{s}<\nu_4
        \end{cases}
    \end{split}
    \end{align}
Here we have defined the total mass and chirp mass,
    \begin{equation}
        \label{eq:mass}
        M\equiv m_1+m_2,\qquad\mathcal{M}\equiv\frac{\qty(m_1m_2)^{3/5}}{M^{1/5}}.
    \end{equation}
Recall that the source-frame frequency $\nu_\mathrm{s}$ is a function of the redshift and the peculiar velocities [as given by Eq.~\eqref{eq:source-frequency}], which means that $\mathcal{S}_i$ implicitly depends on these as well.
The frequencies $\nu_1,\nu_2,\nu_3,\nu_4$ are numerical constants for each system, given by
    \begin{equation}
        \nu_i=\frac{1}{\uppi GM}\qty[\nu_i^{(0)}\qty(\chi)+\sum_{j=1}^3\sum_{k=0}^{3-j}y_i^{(jk)}\qty(\frac{\mathcal{M}}{M})^{5j/3}\chi^k],
    \end{equation}
    where the spin parameter $\chi$ is defined as
    \begin{equation}
        \chi\equiv\frac{m_1}{M}\chi_1+\frac{m_2}{M}\chi_2.
    \end{equation}
For the BNS and BHNS cases, we truncate the waveform at the merger frequency $\nu_1$ and ignore any higher frequencies, since the merger and ringdown phases of the above expression are only valid for BBH.
(BNS mergers occur at frequencies above the LIGO-Virgo band anyway, so for observational purposes this has little effect. However, it should be possible to extend our model to higher frequencies by using waveforms that account for neutron-star (NS) matter effects. This may be desirable as future detectors improve our sensitivity to GWs at these frequencies.)
The spectrum is therefore determined for all types of binaries by the numerical constants $\alpha_i$, $\epsilon_i$, $y_i^{(jk)}$ (which are derived from fits between numerical simulations and post-Newtonian expansions), and $c_i$ (which are chosen to ensure the waveform is continuous).
The values used for each of these match those in Refs.~\cite{Ajith:2007kx,Ajith:2009bn}, and are given in Appendix~\ref{sec:waveforms}.

For the binary merger rate per galaxy, $R_i$, we have
    \begin{equation}
        R_i\qty(z,\vb*\zeta_\mathrm{g},\vb*\zeta_\mathrm{b})=\epsilon f_\mathcal{Z}p_i\qty(\vb*\zeta_\mathrm{b})\psi_{\mathrm{d},i},
    \end{equation}
where $p_i(\vb*\zeta_\mathrm{b})=p_i\qty(m_1,m_2,\chi_1,\chi_2)$ is the joint probability distribution for the masses and spins for a compact binary of type $i$.
The fact that not all stars ultimately end up in compact binaries is accounted for by multiplying the delayed SFR by the efficiency factor $\epsilon$ (which is some unknown constant), as well as the factor $f_\mathcal{Z}$, which accounts for the suppression of massive BH formation in high-metallicity environments.
The factor $f_\mathcal{Z}$ enforces the assumption that black holes with mass greater than $30M_\odot$ can only be formed in galaxies with
    \begin{equation}
        Z\le\frac{1}{2}Z_\odot\qquad\Longrightarrow\qquad\mathcal{Z}\le\log_{10}\frac{1}{2}\approx-0.301.
    \end{equation}
Hence one considers a rate correction factor,
    \begin{equation}
        f_\mathcal{Z}\qty(\mathcal{Z},m_1,m_2)=
        \begin{cases}
            1, & m_1,m_2<30M_\odot\\
            \Theta\qty(\log_{10}\frac{1}{2}-\mathcal{Z}), & \text{else}
        \end{cases}
    \end{equation}
    where $\Theta\qty(x)$ is the Heaviside step function.
The delay time distribution is modeled as $p\qty(t_\mathrm{d})\propto1/t_\mathrm{d}$ between some minimum delay time $t_{\mathrm{min},i}$ (20 Myr for BNS, 50 Myr for BBH and BHNS) and the maximum $t_\mathrm{max}$ equal to the age of the Universe at redshift $z$, namely
    \begin{equation}
        \label{eq:age}
        t\qty(z)=\frac{2t_H}{3\Omega_\Lambda^{1/2}}\mathrm{arcsinh}\qty(\sqrt{\frac{\Omega_\Lambda}{\Omega_\mathrm{m}\qty(1+z)^3}}).
    \end{equation}
At redshift zero this reduces to $t_\mathrm{o}\approx0.958\,t_H$, the current age of the Universe.
The delayed SFR is therefore
    \begin{equation}
        \psi_{\mathrm{d},i}=\frac{1}{\ln\qty(t\qty(z)/t_{\mathrm{min},i})}\int_{t_{\mathrm{min},i}}^{t\qty(z)}\dd{\qty(\ln t_\mathrm{d})}\psi\qty(z_\mathrm{f}),
    \end{equation}
    where the prefactor gives the appropriate normalization of the probability distribution.

It remains to specify the source parameter distributions $p_i\qty(\vb*\zeta_\mathrm{b})$.
We take neutron stars as having masses uniformly distributed between $1M_\odot$ and $2M_\odot$, with zero spin.
The latter is motivated by pulsar observations and by parameter estimation on GW170817, which both indicate that BNS systems typically have low spins ($\chi\lesssim0.05$) by the time they merge~\cite{Abbott:2018wiz} (in any case, we find that spin has little effect on the final results).
For black holes, the primary mass is given by a Salpeter initial mass function $\propto m_1^{-2.35}$ between $5M_\odot$ and $95M_\odot$.
If the secondary mass is a black hole, then it is distributed $\propto1/\qty(m_1-5M_\odot)$.
We require the sum of the masses to be less than or equal to $100M_\odot$, and take the black hole spins as uniform between $-1$ and $+1$.
This information can be summarized by writing
\begin{widetext}
    \begin{align}
        \begin{split}
            p_\mathrm{BNS}&\propto\delta\qty(\chi_1)\delta\qty(\chi_2),\\
            p_\mathrm{BBH}&\propto m_1^{-2.35}/\qty(m_1-5M_\odot),\\
            p_\mathrm{BHNS}&\propto m_1^{-2.35}\delta\qty(\chi_2),
        \end{split}
        \begin{split}
            1M_\odot&\le m_1\le2M_\odot,\\
            5M_\odot&\le m_2\le m_1\le95M_\odot,\\
            5M_\odot&\le m_1\le95M_\odot,
        \end{split}
        \quad
        \begin{split}
            1M_\odot\le m_2&\le2M_\odot,\\
            m_1+m_2&\le100M_\odot,\\
            1M_\odot\le m_2&\le2M_\odot,
        \end{split}
    \end{align}
\end{widetext}
    with a normalising constant chosen appropriately in each case such that $\int\dd{\vb*\zeta_\mathrm{b}}p_i\qty(\vb*\zeta_\mathrm{b})=1$.

The rate $R_i$ is not yet fully determined, as the constant $\epsilon$ is unknown.
However, we can eliminate this constant using the \emph{local} (i.e. redshift zero) rates inferred by LIGO/Virgo.
We do this by calculating the total rate of mergers of type $i$ per unit comoving volume at redshift $z$,
    \begin{align}
    \begin{split}
        \mathcal{R}_i\qty(z)&\equiv\int\dd{\vb*\zeta_\mathrm{g}}\bar{n}\int\dd{\vb*\zeta_\mathrm{b}}R_i\qty(z,\vb*\zeta_\mathrm{g},\vb*\zeta_\mathrm{b})\\
        &=\epsilon\int\dd{\vb*\zeta_\mathrm{g}}\bar{n}\int\dd{\vb*\zeta_\mathrm{b}} f_\mathcal{Z}p_i\qty(\vb*\zeta_\mathrm{b})\psi_{\mathrm{d},i}.
    \end{split}
    \end{align}
We require that in the limit $z\to0$ this matches the local rates inferred by LIGO/Virgo, so $\mathcal{R}_i\qty(0)=\mathcal{R}_i^{(\mathrm{local})}$.
The most up-to-date values are~\cite{Abbott:2017xzg}
    \begin{align}
    \begin{split}
        \mathcal{R}_\mathrm{BNS}^{(\mathrm{local})}&=1.54\times10^{-6}\text{ Mpc}^{-3}\text{yr}^{-1},\\
        \mathcal{R}_\mathrm{BBH}^{(\mathrm{local})}&=1.03\times10^{-7}\text{ Mpc}^{-3}\text{yr}^{-1}.
    \end{split}
    \end{align}
LIGO and Virgo have not yet observed any BHNS events, so it is only possible to place an upper limit of \cite{Abbott:2016ymx}
    \begin{equation}
        \label{eq:BHNS-upper-limit}
        \mathcal{R}_\mathrm{BHNS}^{(\mathrm{local})}\le3.60\times10^{-6}\text{ Mpc}^{-3}\text{yr}^{-1}.
    \end{equation}
Based on previous population synthesis studies, the true BHNS rate is expected to be somewhere between the BNS and BBH rates (e.g., see Ref.~\cite{Abadie:2010cf} for a review), and therefore roughly an order magnitude less than this upper limit.
In order to remain agnostic about the true value of the BHNS rate, we consider two cases: one where we include BHNS mergers at the maximal rate Eq.~\eqref{eq:BHNS-upper-limit}, and one where we set the BHNS rate to 0.

Matching to these local rates allows us to eliminate $\epsilon$ to find
    \begin{equation}
        R_i=\frac{\mathcal{R}^{(\mathrm{local})}_i}{\mathcal{I}_i}p_i\qty(\vb*\zeta_\mathrm{b})f_\mathcal{Z}\psi_{\mathrm{d},i},
    \end{equation}
    where we have defined the normalizing constants
    \begin{equation}
        \mathcal{I}_i=\int\dd{\vb*\zeta_\mathrm{g}}\left.\qty(\bar{n}\psi_{\mathrm{d},i})\right|_{z=0}\int\dd{\vb*\zeta_\mathrm{b}}p_i\qty(\vb*\zeta_\mathrm{b})f_\mathcal{Z}.
    \end{equation}
This is particularly simple in the BNS case, since $f_\mathcal{Z}=1$, so $\mathcal{I}_\mathrm{BNS}=\int\dd{\vb*\zeta_\mathrm{g}}\left.\qty(\bar{n}\psi_{\mathrm{d},i})\right|_{z=0}$.

Our expression for the SGWB density parameter therefore becomes
    \begin{align}
    \begin{split}
        \Omega_\mathrm{gw}=&\sum_i\frac{\uppi\mathcal{R}^{(\mathrm{local})}_i}{3\mathcal{I}_i}\qty(t_H\nu_\mathrm{o})^3\int_0^{z_\mathrm{max}}\dd{z}\frac{1+z}{E\qty(z)}\\
        &\times\int\dd{\vb*\zeta_\mathrm{g}}\psi_{\mathrm{d},i}\bar{n}\qty(1+\delta_n+\vu*e_\mathrm{o}\vdot\vb*v_\mathrm{o})\int\dd{\vb*\zeta_\mathrm{b}}p_i\qty(\vb*\zeta_\mathrm{b})f_\mathcal{Z}\mathcal{S}_i.
    \end{split}
    \end{align}

\subsection{Decomposing the background}
We can decompose the density parameter $\Omega_\mathrm{gw}$ as
    \begin{equation}
        \Omega_\mathrm{gw}\equiv\bar{\Omega}_\mathrm{gw}\qty(1+\delta_\mathrm{gw}),
    \end{equation}
    where $\bar{\Omega}_\mathrm{gw}$ is the average (isotropic) value of the density parameter over the sky, and $\delta_\mathrm{gw}$ is the GW density contrast, which encodes the anisotropies.
The latter can itself be decomposed as
    \begin{equation}
        \delta_\mathrm{gw}\equiv\delta_\mathrm{gw}^{(\mathrm{s})}+\mathcal{D}\,\vu*e_\mathrm{o}\vdot\vu*v_\mathrm{o},
    \end{equation}
    where $\delta_\mathrm{gw}^{(\mathrm{s})}$ is the density contrast due to the true cosmological anisotropies, and the latter term is the kinematic dipole, with direction $\vu*v_\mathrm{o}\equiv\vb*v_\mathrm{o}/\qty|\vb*v_\mathrm{o}|$ and magnitude given by the dipole factor $\mathcal{D}$.
This factor can be calculated by performing a Taylor expansion in $x\equiv1+\vu*e_\mathrm{o}\vdot\vb*v_\mathrm{o}$ around $x=1$, giving
    \begin{equation}
        \mathcal{D}\qty(\nu_\mathrm{o})\equiv v_\mathrm{o}\bar{\Omega}_\mathrm{gw}^{-1}\left.\pdv{\Omega_\mathrm{gw}}{x}\right|_{x=1,\delta_n=0},
    \end{equation}
    where $v_\mathrm{o}\equiv\qty|\vb*v_\mathrm{o}|$.
For our particular case, we therefore have
\begin{widetext}
    \begin{align}
        \bar{\Omega}_\mathrm{gw}&=\sum_i\frac{\uppi\mathcal{R}^{(\mathrm{local})}_i}{3\mathcal{I}_i}\qty(t_H\nu_\mathrm{o})^3\int_0^{z_\mathrm{max}}\dd{z}\frac{1+z}{E\qty(z)}\int\dd{\vb*\zeta_\mathrm{g}}\psi_{\mathrm{d},i}\bar{n}\int\dd{\vb*\zeta_\mathrm{b}}p_i\qty(\vb*\zeta_\mathrm{b})f_\mathcal{Z}\mathcal{S}_i,\\
        \delta_\mathrm{gw}^{(\mathrm{s})}&=\bar{\Omega}_\mathrm{gw}^{-1}\sum_i\frac{\uppi\mathcal{R}^{(\mathrm{local})}_i}{3\mathcal{I}_i}\qty(t_H\nu_\mathrm{o})^3\int_0^{z_\mathrm{max}}\dd{z}\frac{1+z}{E\qty(z)}\int\dd{\vb*\zeta_\mathrm{g}}\psi_{\mathrm{d},i}\bar{n}\delta_n\int\dd{\vb*\zeta_\mathrm{b}}p_i\qty(\vb*\zeta_\mathrm{b})f_\mathcal{Z}\mathcal{S}_i,\\
        \mathcal{D}&=v_\mathrm{o}\bar{\Omega}_\mathrm{gw}^{-1}\sum_i\frac{\uppi\mathcal{R}^{(\mathrm{local})}_i}{3\mathcal{I}_i}\qty(t_H\nu_\mathrm{o})^3\int_0^{z_\mathrm{max}}\dd{z}\frac{1+z}{E\qty(z)}\int\dd{\vb*\zeta_\mathrm{g}}\psi_{\mathrm{d},i}\bar{n}\int\dd{\vb*\zeta_\mathrm{b}}p_i\qty(\vb*\zeta_\mathrm{b})f_\mathcal{Z}\pdv{}{x}\qty(x\mathcal{S}_i).
    \end{align}
\end{widetext}

The anisotropies are then characterized by the overdensity field $\delta^{(\mathrm{s})}_\mathrm{gw}$, either directly or in terms of its statistics.
One particularly useful statistical descriptor is the two-point correlation function (2PCF), defined as the second moment of the overdensity field,
    \begin{equation}
        C_\mathrm{gw}\qty(\theta_\mathrm{o},\nu_\mathrm{o})\equiv\ev{\delta_\mathrm{gw}^\qty(\mathrm{s})\qty(\nu_\mathrm{o},\vu*e_\mathrm{o})\delta_\mathrm{gw}^\qty(\mathrm{s})(\nu_\mathrm{o},\vu*e\mathrlap{'}_\mathrm{o})},
    \end{equation}
    where $\theta_\mathrm{o}\equiv\cos^{-1}(\vu*e_\mathrm{o}\vdot\vu*e\mathrlap{'}_\mathrm{o})$, and the angle brackets denote an averaging over all pairs of directions $\vu*e_\mathrm{o}$, $\vu*e\mathrlap{'}_\mathrm{o}$ whose angle of separation is $\theta_\mathrm{o}$, as well as an ensemble averaging over possible random realizations of the SGWB.
The first moment (i.e., mean) vanishes by definition, and if the background is a Gaussian random field (GRF) then all higher moments either vanish or are expressed in terms of the second moment by Wick's theorem.
The 2PCF therefore uniquely characterizes the anisotropies in the Gaussian part of the background.
It is common practice (particularly in the CMB literature) to perform a multipole expansion of the 2PCF,
    \begin{equation}
        \label{eq:multipole-expansion}
        C_\mathrm{gw}\qty(\theta_\mathrm{o},\nu_\mathrm{o})=\sum_{\ell=0}^\infty\frac{2\ell+1}{4\uppi}C_\ell\qty(\nu_\mathrm{o})P_\ell(\cos\theta_\mathrm{o}),
    \end{equation}
    where $P_\ell\qty(x)$ denotes the $\ell\mathrm{th}$ Legendre polynomial.
The anisotropies are then described in terms of the $C_\ell$ components, which are given by
    \begin{equation}
        C_\ell\qty(\nu_\mathrm{o})\equiv2\uppi\int_{-1}^{+1}\dd{\qty(\cos\theta_\mathrm{o})}C_\mathrm{gw}\qty(\theta_\mathrm{o},\nu_\mathrm{o})P_\ell\qty(\cos\theta_\mathrm{o}).
    \end{equation}
The quantity $\ell\qty(\ell+1)C_\ell/2\uppi$ is roughly the contribution to the anisotropic variance of $\delta^{(\mathrm{s})}_\mathrm{gw}$ per logarithmic bin in $\ell$, as can be seen by considering
    \begin{align}
    \begin{split}
        \mathrm{var}\qty(\delta^{(\mathrm{s})}_\mathrm{gw})&=C_\mathrm{gw}\qty(\theta_\mathrm{o}=0)=\sum_\ell\frac{2\ell+1}{4\uppi}C_\ell\\
        &\approx\int\dd{\qty(\ln\ell)}\frac{\ell\qty(\ell+1)C_\ell}{2\uppi}.
    \end{split}
    \end{align}

While the overdensity $\delta^{(\mathrm{s})}_\mathrm{gw}$ is a random field, the $C_\ell$'s are treated as deterministic quantities, averaged over some ensemble of realizations of the SGWB (e.g., as observed at different locations in the Universe).
As a result, any measurement of the $C_\ell$'s has some cosmological uncertainty associated with it, due to the fact that we can only access a single realization of the SGWB.
This is called the cosmic variance, and is given by
    \begin{equation}
        \mathrm{var}\qty(C_\ell)=\frac{2}{2\ell+1}C_\ell^2.
    \end{equation}

\subsection{Non-Gaussianity in the AGWB}
\label{sec:non-Gaussian}
As mentioned above, the analysis of the SGWB is simplified if it is a GRF, as this eliminates the need for anything other than the second moment of the overdensity (i.e. the 2PCF).
In Ref.~\cite{Jenkins:2018nty}, it was found that a GW background composed of independent sources and discretized into $N_\mathrm{pix}$ pixels on the sky is a GRF at frequency $\nu$ if it satisfies
    \begin{equation}
        \nu T\gg\frac{N_\mathrm{pix}}{\Lambda}+\frac{N_\mathrm{pix}^2}{\Lambda^2},
    \end{equation}
    where $T$ is the observation time and $\Lambda$ is the average number of signals in-band at any given moment (this is referred to as the duty cycle).
However, this was derived by assuming that the duration of a signal at frequency $\nu$ is roughly $1/\nu$, which is a good approximation for burstlike signals, but is inaccurate for the chirp signals emitted by coalescing compact binaries.
As mentioned in Ref.~\cite{Jenkins:2018nty}, the duration of a compact binary signal in some small frequency interval $[\nu,\nu+\updelta\nu]$ is given by
    \begin{equation}
      \upDelta t\approx\qty(\frac{96}{5}\uppi^{8/3}\mathcal{M}^{5/3}\nu^{11/3})^{-1}\updelta\nu,
    \end{equation}
    where $\mathcal{M}$ is the chirp mass of the binary, as defined in Eq.~\eqref{eq:mass}.
Taking this into account, the appropriate limit on the observing time becomes
    \begin{equation}
        T\gg\frac{N_\mathrm{pix}}{R}+\frac{N_\mathrm{pix}^2}{R^2\upDelta t},
    \end{equation}
    where $R$ is the average rate of arrival of GW signals.

If we focus on the AGWB at a single frequency, then we are choosing a frequency interval $\updelta\nu$ equal to the frequency resolution $1/T$.
The above implies that the background is only a GRF at this frequency if
    \begin{equation}
        \frac{96}{5}\uppi^{8/3}\mathcal{M}^{5/3}\nu^{11/3}\frac{N_\mathrm{pix}^2}{R^2}\ll1,
    \end{equation}
    which is impossible for the sources considered here.
It is therefore inconsistent to assume Gaussianity when considering the AGWB at a single frequency.
(We note in passing that due to the large exponent on the frequency in the equation above, this requirement is much easier to fulfil in the LISA frequency band.)

There are two possible ways of addressing this: (i) we can choose a larger frequency interval, integrate the signal over this interval, and treat the result as a GRF; (ii) we can characterize the anisotropies at a single frequency by computing higher-order correlators as well as the 2PCF (e.g., the AGWB bispectrum and trispectrum).
The former is computationally very expensive, due to the required sampling of many points in GW frequency space (however, it is worth noting that LIGO/Virgo stochastic searches typically integrate the data over a frequency bin that is significantly larger than $1/T$ anyway, so it may desirable to compute predictions for this approach regardless of considerations about Gaussianity).
The latter requires a more detailed study, but has the potential to reveal more detailed astrophysical and cosmological information.
This will be investigated in a future work.

For the purposes of this work, we focus on the 2PCF computed at a single frequency, as this still contains a great deal of important information, and is the first step in either of the approaches described above.

\section{Analytical approach}
\label{sec:analytical}
In this section, we use simple analytical functions for the galaxy number density $\bar{n}$ and galaxy-galaxy 2PCF $\ev{\delta_n\delta_n}$.
This allows us to derive a simple closed-form expression for each of the $C_\ell$'s, up to a frequency-dependent factor $\mathcal{A}_\mathrm{gw}$ which can be integrated numerically.

\subsection{Galaxy distribution}
In Sec.~\ref{sec:AGWB} we have not yet specified the mean number density distribution $\bar{n}$.
For this initial analytical approach, we assume that the two galaxy parameters are independent---i.e., that there is no correlation between the metallicity of a galaxy and its delayed SFR.
This allows us to write
    \begin{equation}
        \bar{n}\qty(z,\psi_\mathrm{d},\mathcal{Z})=\bar{N}\qty(z)p\qty(\psi_\mathrm{d}|z)p\qty(\mathcal{Z}|z),
    \end{equation}
    where $p\qty(\psi_\mathrm{d}|z)$ and $p\qty(\mathcal{Z}|z)$ are redshift-dependent probability distributions for the two parameters, and
    \begin{equation}
        \bar{N}\qty(z)\equiv\int\dd{\vb*\zeta_\mathrm{g}}\bar{n}
    \end{equation}
    is the total number density of galaxies per comoving volume.
With this assumption, the density parameter $\Omega_\mathrm{gw}$ reads
    \begin{align}
    \begin{split}
        \label{eq:omega-analytical}
        \Omega_\mathrm{gw}=&\sum_i\frac{\uppi\mathcal{R}^{(\mathrm{local})}_i}{3\mathcal{I}_i}\qty(t_H\nu_\mathrm{o})^3\\
        &\times\int_0^{z_\mathrm{max}}\dd{z}\frac{1+z}{E\qty(z)}\bar{N}\qty(z)\qty(\int_0^\infty\dd{\psi_{\mathrm{d},i}}p\qty(\psi_{\mathrm{d},i}|z)\psi_{\mathrm{d},i})\\
        &\times\int_{-\infty}^{\mathcal{Z}_\mathrm{max}}\dd{\mathcal{Z}}p\qty(\mathcal{Z}|z)\int\dd{\vb*\zeta_\mathrm{b}}p_i\qty(\vb*\zeta_\mathrm{b})f_\mathcal{Z}\mathcal{S}_i\qty(1+\delta_n+\vu*e_\mathrm{o}\vdot\vb*v_\mathrm{o}).
    \end{split}
    \end{align}
Notice that since the integrand is proportional to $\psi_\mathrm{d}$, we only need the first moment (i.e. the mean) of the corresponding probability distribution,
    \begin{equation}
        \label{eq:mean-psi}
        \bar{\psi}_\mathrm{d}\equiv\int_0^\infty\dd{\psi_\mathrm{d}}p\qty(\psi_\mathrm{d}|z)\psi_\mathrm{d},
    \end{equation}
    and there is no need to specify anything else about the distribution.
The quantity in Eq.~\eqref{eq:mean-psi} is just the mean delayed SFR, averaged across all galaxies at the appropriate redshifts.
Equivalently, we can think of this as the delayed form of the cosmic mean SFR per galaxy $\bar{\psi}$,
    \begin{equation}
        \bar{\psi}_\mathrm{d}=\frac{1}{\ln\qty(t\qty(z)/t_\mathrm{min})}\int_{t_\mathrm{min}}^{t\qty(z)}\dd{\qty(\ln t_\mathrm{d})}\bar{\psi}\qty(z_\mathrm{f}).
    \end{equation}
Expressions for the cosmic mean SFR are more appropriately given in units per unit comoving volume rather than per galaxy, so we define
    \begin{equation}
        \bar{\psi}^{(V)}\qty(z)\equiv\bar{N}\qty(z)\bar{\psi}\qty(z),
    \end{equation}
    with the mean number of galaxies per comoving volume $\bar{N}$ converting between the two.
This function is given in Ref.~\cite{Vangioni:2014axa} by
    \begin{equation}
        \label{eq:psi}
        \bar{\psi}^{(V)}\qty(z)=\bar{\psi}^{(V)}_\mathrm{peak}\frac{\alpha\exp\qty[\beta\qty(z-z_\mathrm{peak})]}{\alpha-\beta+\beta\exp\qty[\alpha\qty(z-z_\mathrm{peak})]},
    \end{equation}
    with dimensionless constants $\alpha=2.80$, $\beta=2.62$, and with the normalization relative to the peak value $\bar{\psi}^{(V)}_\mathrm{peak}=0.145\,M_\odot\,\text{yr}^{-1}\,\text{Mpc}^{-3}$ at redshift $z_\mathrm{peak}=1.86$.
We thus write the mean of the delayed SFR distribution as
    \begin{equation}
        \bar{\psi}_\mathrm{d}=\frac{1}{\ln\qty(t\qty(z)/t_\mathrm{min})}\int_{t_\mathrm{min}}^{t\qty(z)}\dd{\qty(\ln t_\mathrm{d})}\frac{\bar{\psi}^{(V)}\qty(z_\mathrm{f})}{\bar{N}\qty(z_\mathrm{f})}.
    \end{equation}
No further information or assumptions about the SFR distribution are needed.
We do, however, need an expression for the total galaxy number density $\bar{N}\qty(z)$, in order to convert the SFR per comoving volume $\bar{\psi}^{(V)}$ back into SFR per galaxy.
We use Ref.~\cite{2016ApJ...830...83C}, where a fit to observational data over a redshift range $0\le z\le8$ has been performed, to get
    \begin{equation}
        \log_{10}\qty(\frac{\bar{N}\qty(z)}{1\text{ Mpc}^{-3}})=-0.26-1.08\log_{10}\qty(\frac{t\qty(z)}{1\text{ Gyr}}),
    \end{equation}
    where $t\qty(z)$ is the age of the Universe at redshift $z$, as given by Eq.~\eqref{eq:age}.

The dependence of the integrand in Eq.~\eqref{eq:omega-analytical} on the metallicity is more complicated due to the factor $f_\mathcal{Z}$; we must therefore specify the full distribution $p\qty(\mathcal{Z}|z)$, and not just its mean.
Following Ref.~\cite{Abbott:2017xzg}, we take the metallicity distribution as a Gaussian with variance $1/4$ centered on the cosmic mean metallicity at that redshift, namely
    \begin{equation}
        \label{eq:metallicity}
        p\qty(\mathcal{Z}|z)\propto\exp\qty[-2\qty(\mathcal{Z}-\bar{\mathcal{Z}}\qty(z))^2].
    \end{equation}
Note that this is not strictly a Gaussian, as $\mathcal{Z}$ has a finite maximum value of $\mathcal{Z}_\mathrm{max}$.
The distribution Eq.~\eqref{eq:metallicity} must be reweighted accordingly to give
    \begin{equation}
    p\qty(\mathcal{Z}|z)=\frac{\sqrt{\frac{8}{\uppi}}\,\exp\qty[-2\qty(\mathcal{Z}-\bar{\mathcal{Z}}\qty(z))^2]}{1-\erf\qty[\sqrt{2}\qty(\bar{\mathcal{Z}}\qty(z)-\mathcal{Z}_\mathrm{max})]}.
    \end{equation}
The cosmic mean metallicity is well modeled by \cite{Belczynski:2016obo}
    \begin{align}
    \begin{split}
        \bar{\mathcal{Z}}\qty(z)&\equiv\log_{10}\frac{\bar{Z}\qty(z)}{Z_\odot}\\
        &=\mathcal{Z}_\mathrm{max}+\log_{10}\qty(\frac{1}{\psi^{(V)}_\mathrm{norm}}\int_z^{z_\mathrm{max}}\dd{z'}\frac{\bar{\psi}^{(V)}\qty(z')}{\qty(1+z')E\qty(z')}),
    \end{split}
    \end{align}
    with $\bar{\psi}^{(V)}$ being the mean SFR from Eq.~\eqref{eq:psi} and $\psi^{(V)}_\mathrm{norm}$ a normalizing constant with units of $M_\odot\,\text{yr}^{-1}\,\text{Mpc}^{-3}$, given by \cite{Belczynski:2016obo}
    \begin{equation}
    \psi^{(V)}_\mathrm{norm}=\frac{\rho_\mathrm{bary}H_\mathrm{o}}{\sqrt{10}\,y\qty(1-R)}=8.90\,M_\odot\,\text{yr}^{-1}\,\text{Mpc}^{-3},
    \end{equation}
    where $\rho_\mathrm{bary}$ is the present-day baryon density, $y=0.019$ is the net metal yield created by each generation of stars, and $R=0.27$ is the fraction of metals returned to the interstellar medium by stars.

Given these distributions, the isotropic GW energy density is therefore
\begin{widetext}
    \begin{align}
    \begin{split}
        \label{eq:omega-analytical_2}
        \bar{\Omega}_\mathrm{gw}=\sum_i\frac{\uppi\mathcal{R}^{(\mathrm{local})}_i}{3\mathcal{I}_i}\qty(t_H\nu_\mathrm{o})^3&\int_0^{z_\mathrm{max}}\frac{\dd{z}\qty(1+z)}{E\qty(z)\ln\qty(t\qty(z)/t_{\mathrm{min},i})}\\
        &\times\qty(\int_{t_{\mathrm{min},i}}^{t\qty(z)}\dd{\qty(\ln t_\mathrm{d})}\bar{\psi}^{(V)}\qty(z_\mathrm{f})\frac{\bar{N}\qty(z)}{\bar{N}\qty(z_\mathrm{f})})\int_{-\infty}^{\mathcal{Z}_\mathrm{max}}\dd{\mathcal{Z}}p\qty(\mathcal{Z}|z)\int\dd{\vb*\zeta_\mathrm{b}}p_i\qty(\vb*\zeta_\mathrm{b})f_\mathcal{Z}\mathcal{S}_i.
    \end{split}
    \end{align}
    with the rate normalizing factor given by
    \begin{equation}
        \mathcal{I}_i=\left.\qty[\frac{1}{\ln\qty(t\qty(z)/t_{\mathrm{min},i})}\qty(\int_{t_{\mathrm{min},i}}^{t\qty(z)}\dd{\qty(\ln t_\mathrm{d})}\bar{\psi}^{(V)}\qty(z_\mathrm{f})\frac{\bar{N}\qty(z)}{\bar{N}\qty(z_\mathrm{f})})\int_{-\infty}^{\mathcal{Z}_\mathrm{max}}\dd{\mathcal{Z}}p\qty(\mathcal{Z}|z)\int\dd{\vb*\zeta_\mathrm{b}}p_i\qty(\vb*\zeta_\mathrm{b})f_\mathcal{Z}]\right|_{z=0}.
    \end{equation}
The result is shown in Fig.~\ref{fig:monopole}.
\begin{figure*}[t]
    \includegraphics[width=\textwidth]{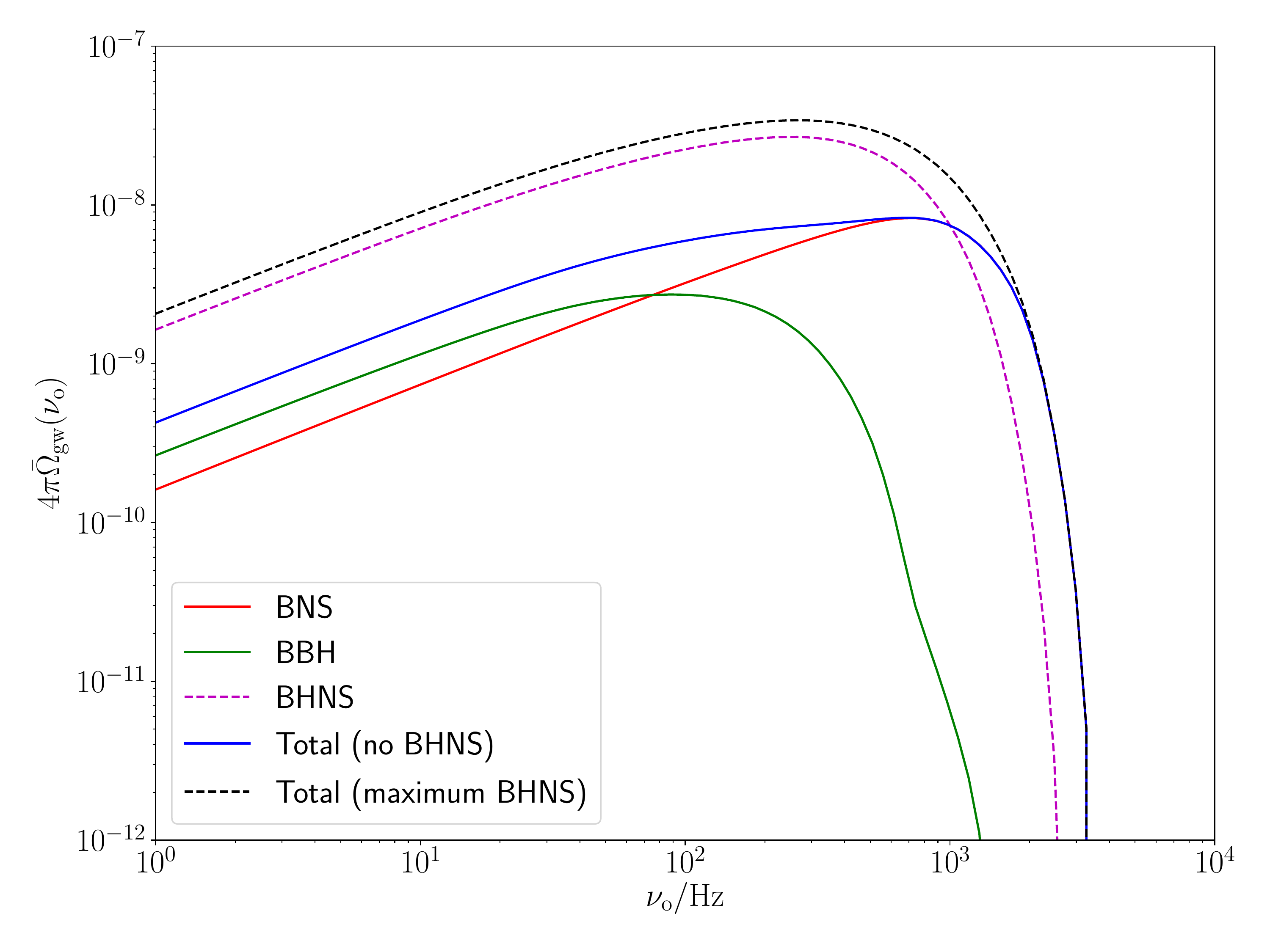}
    \caption{The total GW flux spectrum $4\uppi\bar{\Omega}_\mathrm{gw}$ given by numerically integrating Eq.~\eqref{eq:omega-analytical_2}. (The factor of $4\uppi$ is included to give the \emph{total} flux, rather than the average flux per solid angle.) The total flux with (without) BHNS mergers is given in black (blue), while the flux coming from BNS, BBH, and BHNS mergers is indicated in red, green, and magenta, respectively. The curves including the BHNS contribution are dashed to indicate that this is an upper bound only.}
    \label{fig:monopole}
\end{figure*}

\subsection{Anisotropies}
We now consider anisotropies in the AGWB.
Since we do not know the form of $\delta_n$ as a function of sky position, we resort to a statistical description.
If the signal is Gaussian, then a complete and simple description is given by the 2PCF.
Even if the signal is non-Gaussian (as suggested by Sec.~\ref{sec:non-Gaussian}), this is still a convenient and informative descriptor of the anisotropies.
We write the 2PCF of the stochastic background as
    \begin{align}
    \begin{split}
        C_\mathrm{gw}\qty(\theta_\mathrm{o},\nu_\mathrm{o})=\frac{\uppi^2\qty(t_H\nu_\mathrm{o})^6}{9\bar{\Omega}_\mathrm{gw}^2}\int_0^{z_\mathrm{max}}&\dd{z}\frac{1+z}{E\qty(z)}\int_0^{z_\mathrm{max}}\dd{z'}\frac{1+z'}{E\qty(z')}\qty(\sum_i\int\dd{\vb*\zeta_\mathrm{g}}\bar{n}\qty(z,\vb*\zeta_\mathrm{g})\int\dd{\vb*\zeta_\mathrm{b}}R_i\qty(z,\vb*\zeta_\mathrm{g},\vb*\zeta_\mathrm{b})\mathcal{S}_i\qty(\nu_\mathrm{s},\vb*\zeta_\mathrm{b}))\\
        &\times\qty(\sum_j\int\dd{\vb*\zeta_\mathrm{g}'}\bar{n}\qty(z',\vb*\zeta_\mathrm{g}')\int\dd{\vb*\zeta_\mathrm{b}'}R_j\qty(z',\vb*\zeta_\mathrm{g}',\vb*\zeta_\mathrm{b}')\mathcal{S}_j\qty(\nu_\mathrm{s}',\vb*\zeta_\mathrm{b}'))\ev{\delta_n\qty(z,\vu*e_\mathrm{o},\vb*\zeta_\mathrm{g})\delta_n\qty(z',\vu*e\mathrlap{'}_\mathrm{o},\vb*\zeta_\mathrm{g}')},
    \end{split}
    \end{align}
\end{widetext}
which is determined by the final factor on the rhs, namely, the galaxy-galaxy 2PCF, $\xi_\mathrm{gg}\equiv\ev{\delta_n\delta_n}$.
There exists a simple analytical estimate for this,
    \begin{equation}
        \label{eq:galaxy-2pcf}
        \xi_\mathrm{gg}\qty(d)\approx\qty(\frac{d}{d_1})^{-\gamma}\delta\qty(z-z'),
    \end{equation}
    where $d\qty(z,\vu*e_\mathrm{o},\vu*e\mathrlap{'}_\mathrm{o})$ is the comoving distance between the two points, $d_1$ is the comoving scale at which the correlation is unity, and $\gamma$ is some positive real number.
We assume for simplicity that all galaxies cluster in the same way, regardless of their SFR and metallicity, and that $d_1$ and $\gamma$ are constant, and take the values\footnote{The quoted values are taken from the VIMOS Public Extragalactic Redshift Survey (VIPERS)~\cite{Marulli:2013wpa}, and are valid for the brightest galaxies in a redshift bin centered on $z=1$, which are precisely the galaxies that contribute most strongly to $\Omega_\mathrm{gw}$.} $d_1=(4.29\pm0.19)h^{-1}\text{Mpc}$ and $\gamma=1.63\pm0.04$.
Writing the comoving distance $d$ as
    \begin{equation}
        d=2r\tan\frac{\theta_\mathrm{o}}{2},\qquad\theta_\mathrm{o}\equiv\cos^{-1}\qty(\vu*e_\mathrm{o}\vdot\vu*e\mathrlap{'}_\mathrm{o}),
    \end{equation}
    where $r=\int\dd{z}/H$ is the comoving distance between the observer and the galaxy, the 2PCF can be simply written as
    \begin{equation}
        \label{eq:C-gw}
        C_\mathrm{gw}\qty(\theta_\mathrm{o},\nu_\mathrm{o})=\mathcal{A}_\mathrm{gw}\qty(\nu_\mathrm{o})\qty(\tan\frac{\theta_\mathrm{o}}{2})^{-\gamma},
    \end{equation}
    where the ``amplitude" of the correlation $\mathcal{A}_\mathrm{gw}$ is a function of the GW frequency alone, and has no angular dependence,
    \begin{equation}
        \label{eq:A-gw}
        \mathcal{A}_\mathrm{gw}\qty(\nu_\mathrm{o})\equiv\bar{\Omega}_\mathrm{gw}^{-2}\qty(\frac{d_1}{2})^\gamma\int_0^{z_\mathrm{max}}\frac{\dd{z}}{r^\gamma}\qty(\pdv{\bar{\Omega}_\mathrm{gw}}{z})^2.
    \end{equation}
The multipole components $C_\ell$ are then given by
    \begin{equation}
        C_\ell\qty(\nu_\mathrm{o})=2\uppi\mathcal{A}_\mathrm{gw}\int_{-1}^{+1}\dd{\qty(\cos\theta_\mathrm{o})}P_\ell\qty(\cos\theta_\mathrm{o})\qty(\tan\frac{\theta_\mathrm{o}}{2})^{-\gamma},
    \end{equation}
which can be evaluated explicitly (see Appendix~\ref{sec:C-ell-analytical}) to give
    \begin{equation}
        \label{eq:C-ell-analytical}
        C_\ell=4\uppi\mathcal{A}_\mathrm{gw}\frac{{}_3F_2\qty(-\ell,\ell+1,1-\frac{\gamma}{2};1,2;1)}{\mathrm{sinc}\qty(\uppi\gamma/2)},
    \end{equation}
    where $\mathrm{sinc}\qty(x)\equiv\sin\qty(x)/x$, and ${}_3F_2$ is a generalized hypergeometric function, which is simple to evaluate numerically.

We emphasize that Eq.~\eqref{eq:C-ell-analytical} is only a simple approximation of the true angular spectrum of the anisotropies.
In reality, the galaxy clustering is more complicated than we have assumed, since it evolves with redshift and varies between different populations of galaxies with different physical attributes.
We also note that we have extended Eq.~\eqref{eq:galaxy-2pcf} beyond its realm of validity by assuming that it holds for all distances $d$.
The galaxy 2PCF drops below this power law for distances smaller than $\approx0.1h^{-1}$~Mpc and tends to some finite value as $d\to0$, and also drops below the power law for distances larger than $\approx10h^{-1}$~Mpc, eventually becoming negative for large separations.
There are two resulting inaccuracies in Eqs.~\eqref{eq:C-gw} and~\eqref{eq:C-ell-analytical} that are immediately evident.
First, the correlation $C_\mathrm{gw}$ diverges as $\theta_\mathrm{o}\to0$, even though this should give a well-defined finite value equal to the variance of the field at each point, $\mathrm{var}(\delta_\mathrm{gw}^{(\mathrm{s})})$.
Secondly, the $\ell=0$ moment of the 2PCF evaluates to $C_0=4\uppi\mathcal{A}_\mathrm{gw}/\mathrm{sinc}\qty(\uppi\gamma/2)>0$, even though $C_0$ should be exactly equal to 0 (as shown in Appendix~\ref{sec:C-ell-analytical}).
These inaccuracies are due to the extrapolation of Eq.~\eqref{eq:galaxy-2pcf} down to $d=0$ and out to $d\to\infty$, respectively.
We therefore turn to a more detailed and accurate approach in the following section, to address the deficiencies of this simple model.

\section{Catalogue approach}
\label{sec:catalogue}
\begin{figure*}[t]
\includegraphics[width=\textwidth]{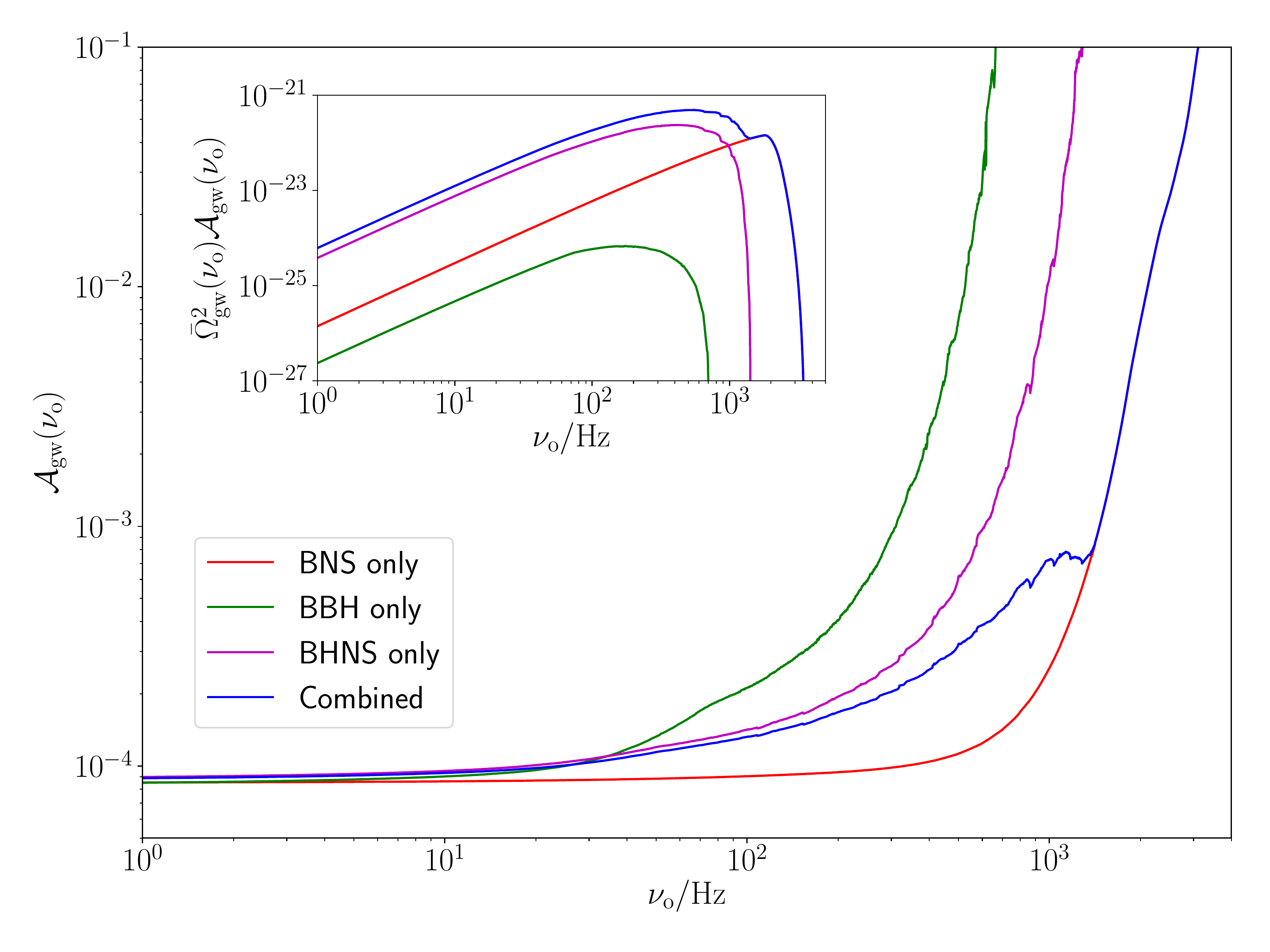}
\caption{The amplitude factor $\mathcal{A}_\mathrm{gw}$ defined in Eq.~\eqref{eq:A-gw}, as a function of GW frequency $\nu_\mathrm{o}$. The lines in red, green, and magenta show what this factor would be if the background included only BNS, only BBH, or only BHNS events, respectively. The blue line shows the combined result, based on the spectra in Fig.~\ref{fig:monopole} (note that the combined curve is not simply the sum of the BBH, BNS and BHNS curves, due to the normalization with respect to $\bar{\Omega}_\mathrm{gw}^2$, which is different in all four cases). The factor diverges at large frequencies because it is normalized with respect to the monopole $\bar{\Omega}_\mathrm{gw}^2$, which vanishes at these frequencies. The enclosed plot shows $\bar{\Omega}_\mathrm{gw}^2\mathcal{A}_\mathrm{gw}$, which encapsulates the \emph{absolute} size of the anisotropies rather than their \emph{relative} size compared to the monopole, and therefore does not diverge.}
\label{fig:A_gw}
\end{figure*}
\begin{figure*}[t]
    \includegraphics[width=\textwidth]{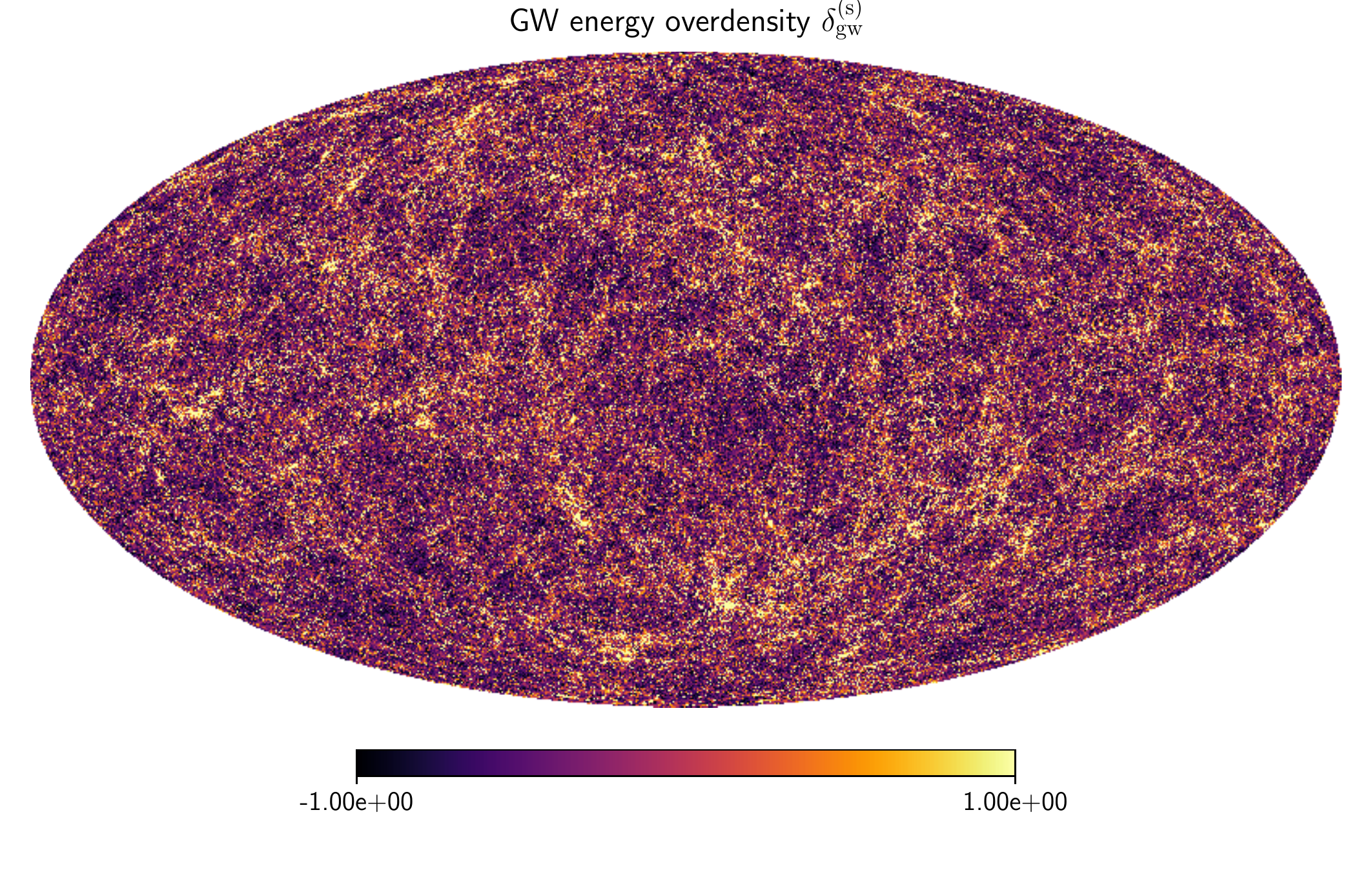}
    \caption{A HEALPix~\cite{Gorski:2004by} map of the GW energy overdensity $\delta_\mathrm{gw}^{(\mathrm{s})}$ constructed from the all-sky mock light cone catalogue~\cite{Blaizot:2003av,DeLucia:2006szx,Springel:2005nw,Lemson:2006ee}, as described in Sec.~\ref{sec:catalogue}. This was generated with the HEALPix $N_\mathrm{side}$ parameter set to 256, corresponding to an angular resolution of 13.7 arcminutes, and an average of 7.3 galaxies per pixel.}
    \label{fig:map}
\end{figure*}
The study of anisotropies in the SGWB induced by the large-scale distribution of astrophysical sources can be also performed using a large enough realistic mock catalogue of galaxies with  recipes to infer the production and merger rates of compact sources. In such catalogues,  the growth of dark matter structure is first simulated and the history of the hierarchical mass assembly is then recorded by means of merger trees of haloes and subhaloes within snapshots stored at different time steps.
To simulate the visible galaxies without performing a cosmological hydrodynamic simulation, one  uses a semianalytic model aiming to reproduce the observed properties of these galaxies (e.g., clustering, counts, scaling relations) at different redshifts.
Such models (see, e.g., Refs.~\cite{Hatton:2003du,DeLucia:2006szx,Henriques:2014sga}) differ by the number of astrophysical processes included to describe the fate of baryons within the dark matter haloes, the phenomenological description of these processes, and their consistency with the various observational data sets.
In that respect, quantitative results strongly depend on the treatment and implementation of feedback mechanisms.
Mock light cones can then be generated from the postprocessed snapshots of the dynamical collisionless N-body simulation, taking care to shuffle them in order to suppress replication effects.

In what follows, we make use of a mock light cone catalogue~\cite{Blaizot:2003av} constructed by applying the ``L-galaxies" model~\cite{DeLucia:2006szx} to the Millennium simulation~\cite{Springel:2005nw,Lemson:2006ee}.
Very relevant for our study are the all-sky coverage and depth of this mock catalogue, even if the random tiling process used to generate a mock light cone (the size of which is larger than the simulated volume) introduces discontinuities in the density field and hence a loss of clustering information.
Hence, the clustering signal suffers from a small negative bias of less than 10\% from 1 to 10 $h^{-1}$~Mpc around the two-point correlation length.
In addition, it exhibits finite volume effects due to the limited size of the simulated box which for instance erases the two-point angular correlation signal at scales larger than a tenth of the simulation box size.
However, both effects are not important for our study since only a fraction of the spatial information is lost affecting scales that are not very relevant for the anisotropies we are looking at.
The comoving box of size $500 h^{-1}$~Mpc on a side of the Millennium simulation has indeed been chosen that large in order to encompass typical scales related to the large-scale features of the spatial distribution of galaxies.

The all-sky mock light cone catalogue \cite{Blaizot:2003av,DeLucia:2006szx} we use contains 5,715,694 galaxies, all limited at apparent AB magnitude of 18 in the r filter from SDSS.
It has been built using a random tiling technique applied to 64 postprocessed snapshots saved during the Millennium simulation with a time step of about 100 Myr.
We queried the database and retrieved for each galaxy the following data: its sky location, cosmological redshift, metallicity, and peculiar velocity.
In order to calculate the delayed star formation rate of each galaxy on the light cone, it was necessary to access information about its star formation rate at earlier snapshots by querying the full Millennium simulation.
This was made more complicated by the fact that the light cone galaxies are the result of a sequence of mergers of smaller galaxies, each with their own independent star formation history.
In order to account for this, we queried the Millennium simulation to extract the full star formation history of each light cone galaxy, given by the star formation rates and redshifts of its progenitors at earlier snapshots.
This included a total of 973,224,532 redshift and SFR measurements from the progenitor galaxies.

We note that due to the magnitude-limited sample used to construct the light cone catalogue, it only extends out to a redshift of $z\approx0.78$.
While this is sufficient to study anisotropies on the angular scales we are interested in, it means that the total energy density $\bar{\Omega}_\mathrm{gw}$ given by the catalogue will be significantly less than the true value, due to the missing contribution from redshifts $z>0.78$.
However, since we describe the anisotropies in terms of the GW overdensity $\delta_\mathrm{gw}^{(\mathrm{s})}$, this will have minimal effect on the $C_\ell$ spectrum.
The main advantage of the mock catalogue lies in its accurate nonlinear modeling of the galaxy clustering statistics, which remains valid on the scales of interest even when the redshift limit is introduced.

In performing our analysis of the mock catalogue, we adhere to the (now outdated) WMAP values of the cosmological parameters that were used in the Millennium simulation: $H_\mathrm{o}=73\text{ km s}^{-1}\text{Mpc}^{-1}$, $\Omega_\mathrm{m}=0.25$, $\Omega_\Lambda=0.75$.
For comparison with the analytical prediction Eq.~\eqref{eq:C-ell-analytical}, we use values of $\gamma$ and $d_1$ that match those of the simulation, which are themselves consistent with the 2-degree Field Galaxy Redshift Survey~\cite{Springel:2005nw,Hawkins:2002sg}: $\gamma=1.67\pm0.03$, $d_1=5.05\pm0.26h^{-1}$~Mpc.
This ensures consistency between our results and the underlying simulation.
We confirm that replacing these with the more up-to-date values mentioned previously ($H_\mathrm{o}=67.9\text{ km s}^{-1}\text{Mpc}^{-1}$, $\Omega_\mathrm{m}=0.3065$, $\Omega_\Lambda=0.6935$, $d_1=4.29h^{-1}\text{ Mpc}$, $\gamma=1.63$) has little impact on the results.

\subsection{Reconstructing the SGWB from pointlike sources}
In order to use the information extracted from the catalogue, we must first explicitly rewrite our equations for $\Omega_\mathrm{gw}$ in terms of the available data for each galaxy.
We do this by expressing the galaxy number density $n$ as a weighted sum of Dirac delta functions, reflecting the fact that we treat each galaxy as a point source.

Let us consider a catalogue of $\mathcal{N}$ galaxies, indexed by a label $k$.
We write their redshift, sky location, delayed star formation rate, and log-normalized metallicity as $z_k$, $\vu*e_k$, $\bar{\psi}_{\mathrm{d},k}\qty(z)$, and $\mathcal{Z}_k$, respectively.
We also allow each galaxy to have a peculiar velocity $v_k$ along the line of sight.
This means that galaxy $k$ has a source-frame frequency given by
    \begin{equation}
        \nu_{\mathrm{s},k}=\nu_\mathrm{o}\qty(1+z_k)\qty(1+v_k-\vu*e_k\vdot\vb*v_\mathrm{o}).
    \end{equation}
By integrating the number density per comoving volume $n$ over redshift, SFR, and metallicity, we must have
    \begin{align}
    \begin{split}
        \mathcal{N}&=\int_{z=0}^{z_\mathrm{max}}\dd[3]{V\qty(z)}\int\dd{\vb*\zeta_\mathrm{g}}n\\
        &=\int_{S^2}\dd[2]{\sigma_\mathrm{o}}\int_0^{z_\mathrm{max}}\dd{z}\frac{r^2}{H}\int_0^\infty\dd{\bar{\psi}_\mathrm{d}}\int_{-\infty}^{\mathcal{Z}_\mathrm{max}}\dd{\mathcal{Z}}n.
    \end{split}
    \end{align}
This implies that
\begin{widetext}
    \begin{equation}
        n=H_\mathrm{o}\sum_k\frac{E\qty(z_k)}{r^2(z_k)}\delta^{(2)}\qty(\vu*e_\mathrm{o},\vu*e_k)\delta\qty(z-z_k)\delta\qty(\bar{\psi}_\mathrm{d}-\bar{\psi}_{\mathrm{d},k}\qty(z_k))\delta\qty(\mathcal{Z}-\mathcal{Z}_k).
    \end{equation}
The SGWB is then given by
    \begin{align}
    \begin{split}
        \Omega_\mathrm{gw}&=\sum_i\frac{\uppi}{3}\qty(t_H\nu_\mathrm{o})^3\int_0^{z_\mathrm{max}}\dd{z}\frac{1+z}{E\qty(z)}\int\dd{\vb*\zeta_\mathrm{g}}n\qty(1+\vu*e_\mathrm{o}\vdot\vb*v_\mathrm{o})\int\dd{\vb*\zeta_\mathrm{b}}R_i\qty(z,\mathcal{Z},\vb*\zeta_\mathrm{b})\mathcal{S}_i\qty(\nu_\mathrm{s},\vb*\zeta_\mathrm{b})\\
        &=\sum_k\sum_i\frac{\uppi H_\mathrm{o}}{3}\qty(t_H\nu_\mathrm{o})^3\frac{1+z_k}{r^2\qty(z_k)}\qty(1+\vu*e_k\vdot\vb*v_\mathrm{o})\int\dd{\vb*\zeta_\mathrm{b}}R_i\qty(z_k,\mathcal{Z}_k,\vb*\zeta_\mathrm{b})\mathcal{S}_i\qty(\nu_{\mathrm{s},k},\vb*\zeta_\mathrm{b})\delta^{(2)}\qty(\vu*e_\mathrm{o},\vu*e_k).
    \end{split}
    \end{align}
Let us remind the reader that $i\in\{\mathrm{BNS},\mathrm{BBH},\mathrm{BHNS}\}$ indexes a sum over the different types of binary merger in each galaxy, while $k\in\{1,2,\ldots,\mathcal{N}\}$ indexes a sum over the galaxies in the catalogue.
The above expression for $\Omega_\mathrm{gw}$ becomes much simpler if we define the weight
    \begin{equation}
        \label{eq:weight}
        w_k\qty(\nu_\mathrm{o})\equiv\sum_i\frac{\uppi H_\mathrm{o}}{3}\qty(t_H\nu_\mathrm{o})^3\frac{1+z_k}{r^2\qty(z_k)}\qty(1+\vu*e_k\vdot\vb*v_\mathrm{o})\int\dd{\vb*\zeta_\mathrm{b}}R_i\qty(z_k,\mathcal{Z}_k,\vb*\zeta_\mathrm{b})\mathcal{S}_i\qty(\nu_{\mathrm{s},k},\vb*\zeta_\mathrm{b})
    \end{equation}
\end{widetext}
    for each galaxy, so that
    \begin{equation}
        \Omega_\mathrm{gw}=\sum_kw_k\,\delta^{(2)}\qty(\vu*e_\mathrm{o},\vu*e_k).
    \end{equation}
In the analytic approach we took in the previous section, we were limited by the fact that $\delta_n$ was not known as a function of sky location; all we knew was the (approximate) 2PCF of $\delta_n$, so we were forced to use the 2PCF of $\Omega_\mathrm{gw}$.
Using a catalogue we now have an explicit expression for $\Omega_\mathrm{gw}$ as a function of sky location, and we can use whatever statistics we like to describe it.

One simple choice is to calculate the spherical multipole components of $\Omega_\mathrm{gw}$,
    \begin{equation}
        \Omega_{\ell m}\qty(\nu_\mathrm{o})\equiv\int_{S^2}\dd[2]{\sigma_\mathrm{o}}\Omega_\mathrm{gw}Y_{\ell m}^*\qty(\vu*e_\mathrm{o})=\sum_kw_kY_{\ell m}^*\qty(\vu*e_k).
    \end{equation}
However, these quantities become expensive to compute if our catalogue contains a large number of galaxies, as each component requires $\mathcal{N}$ evaluations of a spherical harmonic function.
This work focuses on the 2PCF of the GW energy overdensity $C_\mathrm{gw}=\ev{\delta_\mathrm{gw}^{(\mathrm{s})}\delta_\mathrm{gw}^{(\mathrm{s})}}$, which captures much of the information contained in the anisotropies.
It is possible to write down an expression for the $C_\ell$ components of this function in terms of the weights $w_k$ and position vectors $\vu*e_k$ of each galaxy; however, these are even more expensive to compute, as they require a sum over all pairs of galaxies which scales as $\mathcal{N}^2-\mathcal{N}$, or more than $3\times10^{13}$ operations for each $C_\ell$.

In practice, it is much more efficient to compute the statistics of the background using HEALPix\footnote{http://healpix.sourceforge.net}~\cite{Gorski:2004by}, a powerful software package for manipulating pixelized maps of the sphere.
We construct a HEALPix map by computing the weight $w_k$ of each galaxy [given by Eq.~\eqref{eq:weight}] and adding this to the pixel corresponding to the galaxy's sky position $\vu*e_k$.
The $C_\ell$'s (or other relevant quantities) can then be calculated directly in a matter of seconds using HEALPix routines.

\section{Results and discussion}
\label{sec:results-discussion}

\begin{figure*}[t]
    \includegraphics[width=\textwidth]{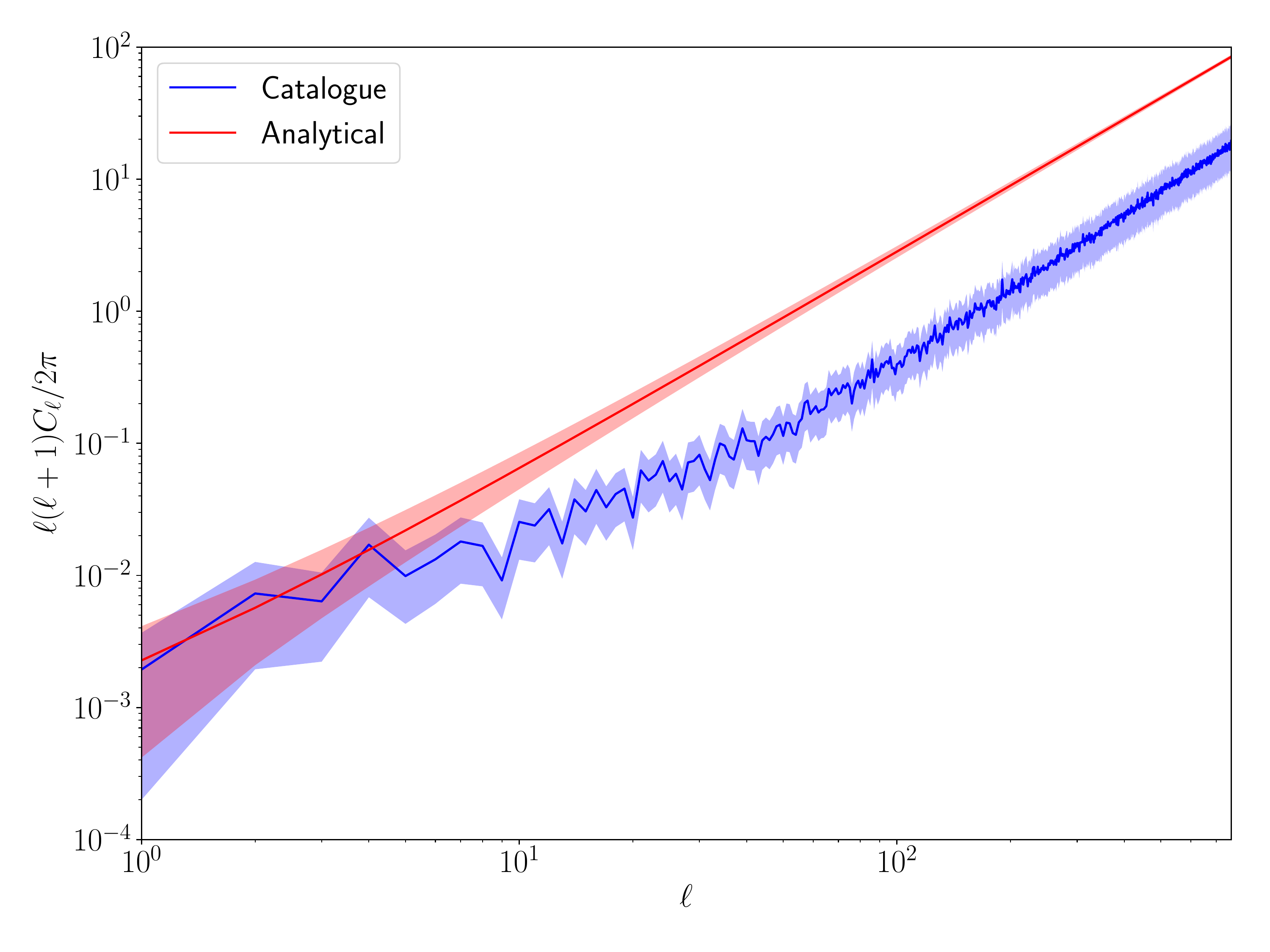}
    \caption{The quantity $\ell\qty(\ell+1)C_\ell/2\uppi$, which is the approximate contribution to the anisotropic variance of $\delta_\mathrm{gw}^{(\mathrm{s})}$ as a function of $\ln\ell$. The red curve shows the simple analytical prediction given in Eq.~\eqref{eq:C-ell-analytical}, while the blue curve shows the spectrum computed from the map in Fig.~\ref{fig:map} using HEALPix. Both curves include error regions from cosmic variance, while the blue curve includes Poisson errors associated with the finite number of galaxies per pixel in the HEALPix map.}
    \label{fig:log-var}
\end{figure*}

We have numerically integrated Eq.~\eqref{eq:omega-analytical_2} to find the frequency spectrum of the isotropic component of the AGWB as shown in Fig.~\ref{fig:monopole}, which matches the corresponding figure in Ref.~\cite{Abbott:2017xzg}.
At frequencies up to $\approx20$~Hz, all sources are included and are in the inspiral regime, giving a monopole $\bar{\Omega}_\mathrm{gw}$ that scales as $\nu_\mathrm{o}^{2/3}$.
The number of emitting sources decreases at higher frequencies, causing the spectrum to taper off and eventually vanish at $\approx3000$~Hz.
Using this spectrum, we are able to compute the factor $\mathcal{A}_\mathrm{gw}$ given in Eq.~\eqref{eq:A-gw}, and hence obtain a simple analytical prediction for the $C_\ell$ spectrum of the background using Eq.~\eqref{eq:C-ell-analytical}.
The value of this factor as a function of frequency is shown in Fig.~\ref{fig:A_gw}.
In the inspiral-dominated regime below $\approx20$~Hz, the value of $\mathcal{A}_\mathrm{gw}$ is essentially constant at $\approx9\times10^{-5}$.
As the frequency increases above this regime, the falloff in the number of sources increases the granularity of the background, and therefore the size of the anisotropies, causing $\mathcal{A}_\mathrm{gw}$ to increase by many orders of magnitude, and eventually diverge as $\bar{\Omega}_\mathrm{gw}\to0$.

Using the above, we can compute the $C_\ell$ at any frequency in the LIGO-Virgo band.
For concreteness, we focus on a frequency of $65.75$~Hz, as this has been used as the reference frequency in previous directional searches for an inspiral-dominated astrophysical background conducted by LIGO/Virgo~\cite{TheLIGOScientific:2016xzw}.
We find
    \begin{equation}
        \mathcal{A}_\mathrm{gw}\qty(\nu_\mathrm{o}=65.75\text{ Hz})\approx1.284\times10^{-4}.
    \end{equation}
This includes BHNS mergers at the maximal rate given in Eq.~\eqref{eq:BHNS-upper-limit}; setting the BHNS rate to 0 gives a 6\% decrease in $\mathcal{A}_\mathrm{gw}$.

We have also computed the $C_\ell$ spectrum of the AGWB from an all-sky mock light cone catalogue based on the Millennium simulation, giving a much more accurate description of the anisotropies by relaxing many of the assumptions about the galaxy-galaxy 2PCF that were used to derive Eq.~\eqref{eq:omega-analytical_2}.
This was done by constructing a GW overdensity map as described in Sec.~\ref{sec:catalogue}, which is shown in Fig.~\ref{fig:map}.
The $C_\ell$ spectrum computed from this map using the HEALPix software package is shown in Fig.~\ref{fig:log-var}, along with the analytical prediction.
Despite the simplicity of the analytical model, and the fact that it contains no free parameters, it is in very strong agreement with the corresponding result from the catalogue at the largest angular scales (i.e., the lowest $\ell$-modes).
It also captures the scaling of $C_\ell$ with $\ell$ at small scales (i.e., large $\ell$), although it overestimates the anisotropic variance at these scales by a constant factor of $\approx4$.
It is worth mentioning that such high values of $\ell$ are inaccessible to the current detector network due to their poor angular resolution---current directional searches only probe the first few $\ell$~\cite{TheLIGOScientific:2016xzw}, although this is expected to improve as more detectors are added to the network.
Note that in order to ensure a fair comparison in Fig.~\ref{fig:log-var}, the analytical curve shown is for the cosmological parameters $H_\mathrm{o}$, $\Omega_\mathrm{m}$, $\Omega_\Lambda$ and galaxy 2PCF parameters $\gamma$, $d_1$ corresponding to the simulation.
Replacing these with the more up-to-date values from Planck~\cite{Ade:2015xua} and VIPERS~\cite{Marulli:2013wpa} has negligible effect.

The results in Fig.~\ref{fig:log-var} can be compared directly with Fig.~2 of Ref.~\cite{Cusin:2018rsq} [with the caveat that Fig.~\ref{fig:log-var} includes BNS and BHNS mergers as well as BBH, whereas Ref.~\cite{Cusin:2018rsq} includes only BBH---however, Fig.~\ref{fig:A_gw} indicates that this should only introduce a factor of $\order{1}$ between the two].
We note that the $C_\ell$'s in Fig.~\ref{fig:log-var} are larger in amplitude than in Ref.~\cite{Cusin:2018rsq}, and do not fall off as quickly with $\ell$.
This perhaps indicates that the linear perturbation-theory approach adopted in Ref.~\cite{Cusin:2018rsq} to describe the galaxy distribution underestimates the clustering at small scales, where the perturbations to the density field become nonlinear.
A more detailed comparison of the two sets of results is needed to fully assess the accuracy and applicability of each approach.

Figure~\ref{fig:log-var} can also be compared with Fig.~5 of Ref.~\cite{Jenkins:2018nty}, which shows the corresponding $C_\ell$ spectra for cosmic string networks with different string tensions $G\mu$.
The anisotropies in the astrophysical background are clearly much larger in amplitude, and become nonlinear for moderate values of $\ell$.
This is unsurprising, given that the bulk of the GW flux from astrophysical sources is emitted at much lower redshifts, so that angular separations on the sky correspond to much smaller physical separations than in the cosmic string case, and therefore to scales where the departure from homogeneity is more evident.

We have also computed the kinematic dipole factor, which in the case shown in Figs.~\ref{fig:map} and~\ref{fig:log-var} has a value of
    \begin{equation}
      \mathcal{D}\qty(\nu_\mathrm{o}=65.75\text{ Hz})\approx7.099\times10^{-4}.
    \end{equation}
This is an order of magnitude smaller than in the cosmic string case, which is intuitively sensible given that the galaxies we consider are generally at much lower redshifts and therefore their velocities due to the Hubble flow are much smaller, giving a weaker Doppler effect.
The fact that $\mathcal{D}$ is smaller in this case, while $\delta_\mathrm{gw}^{(\mathrm{s})}$ is simultaneously larger, means that the effects of the kinematic dipole are unimportant for the astrophysical background.

\section{Conclusion}
We have developed a detailed anisotropic model for the AGWB, including the most important sources for the LIGO-Virgo frequency band (BBH, BNS, and BHNS).
The angular spectrum of the anisotropies, quantified by the $C_\ell$ components, has been calculated through two complementary approaches: a simple, closed-form analytical expression Eq.~\eqref{eq:C-ell-analytical}, and a detailed numerical study using an all-sky mock light cone galaxy catalogue from the Millennium simulation~\cite{Blaizot:2003av,Lemson:2006ee,Springel:2005nw,DeLucia:2006szx}.
The two approaches are in excellent agreement at large angular scales ($\ell\lesssim10$), and differ only by a factor of order unity at smaller scales, following the variation of $\ell\qty(\ell+1)C_\ell/2\uppi$ over many orders of magnitude despite the simplicity of the analytical model and the lack of free parameters.
These anisotropies are considerably larger in amplitude than those in the temperature of the CMB, or those in the SGWB due to cosmic strings~\cite{Jenkins:2018nty}, and become nonlinear at higher multipoles $\ell$.
This shows that modeling the AGWB in a purely isotropic manner neglects a great deal of astrophysical and cosmological information, thereby motivating future theoretical and observational work.
We expect that in the near future, anisotropies in the AGWB (and the SGWB more generally) will become an important probe of the large-scale structure of the Universe, and of the astrophysical processes that occur within it.

We have highlighted several key avenues for future work to explore.
One of the key differences between the SGWB and the CMB is the very broad frequency spectrum over which the SGWB can be investigated.
This motivates the development of models that are valid at frequencies outside of the LIGO-Virgo band; the study of AGWB anisotropies in the LISA band will be of particular interest in the coming decades.
As shown in Sec.~\ref{sec:non-Gaussian}, the use of narrow frequency bins makes it impossible to guarantee that the AGWB is Gaussian, and a full characterization of the anisotropies in this bin requires higher-order correlators of the field (such as the bispectrum and trispectrum).
This is a daunting task, both theoretically and observationally, but it holds the promise of incredibly rich new astrophysical and cosmological information.

There will also be a need to develop increasingly accurate models of the AGWB anisotropies.
This will be addressed in future work by improving upon the fiducial astrophysical model of Ref.~\cite{Abbott:2017xzg}, using past and future LIGO-Virgo observing runs to enhance our knowledge of the relevant sources.
The catalogue approach described in Sec.~\ref{sec:catalogue} will also be improved upon by using more accurate and complete galaxy catalogues.
It will also be interesting to explore different ways in which the statistical properties of the AGWB might differ from those of the model described here; e.g., the extension to models which are statistically nonstationary, or which are no longer statistically isotropic due to the inclusion of sources in the galactic plane of the Milky Way.

Finally, we plan to produce realistic mock data from this model and from future models, and use them to test and optimize data analysis and parameter estimation methods, leading the way to future detections and measurement of AGWB anisotropies.

\begin{acknowledgments}
    We thank Joe Romano and Andrew Matas for reading the manuscript carefully and providing us with valuable comments.
    The Millennium Simulation databases used in this paper and the web application providing online access to them were constructed as part of the activities of the German Astrophysical Virtual Observatory.
    Some of the results in this paper have been derived using the HEALPix package~\cite{Gorski:2004by}.
    A.C.J. is supported by King's College London through a Graduate Teaching Scholarship.
    M.S. is supported in part by the Science and Technology Facility Council (STFC), United Kingdom, under Grant No. ST/P000258/1.
\end{acknowledgments}

\appendix
\section{Explicitly evaluating the analytical \texorpdfstring{$C_\ell$}{} spectrum}
\label{sec:C-ell-analytical}
The $C_\ell$'s are given by
    \begin{equation}
        C_\ell\qty(\nu_\mathrm{o})=2\uppi\mathcal{A}_\mathrm{gw}\int_{-1}^{+1}\dd{\qty(\cos\theta_\mathrm{o})}P_\ell\qty(\cos\theta_\mathrm{o})\qty(\tan\frac{\theta_\mathrm{o}}{2})^{-\gamma}.
    \end{equation}
Applying simple trigonometry, we have
    \begin{equation*}
        \tan\frac{\theta_\mathrm{o}}{2}=\sqrt{\frac{1-\cos\theta_\mathrm{o}}{1+\cos\theta_\mathrm{o}}},
    \end{equation*}
    and therefore
    \begin{equation*}
        C_\ell=2\uppi\mathcal{A}_\mathrm{gw}\int_{-1}^{+1}\dd{x}P_\ell\qty(x)\qty(\frac{1+x}{1-x})^{\gamma/2}.
    \end{equation*}
This can be evaluated by writing the Legendre polynomial as a sum,
    \begin{equation*}
        P_\ell\qty(x)=\sum_{k=0}^\ell\frac{\qty(\ell+k)!}{\qty(k!)^2\qty(\ell-k)!}\qty(\frac{x-1}{2})^k,
    \end{equation*}
    so that we have
    \begin{equation*}
        C_\ell=2\uppi\mathcal{A}_\mathrm{gw}\sum_{k=0}^\ell\frac{\qty(\ell+k)!}{\qty(k!)^2\qty(\ell-k)!}\int_{-1}^{+1}\dd{x}\qty(\frac{1+x}{1-x})^{\gamma/2}\qty(\frac{x-1}{2})^k.
    \end{equation*}
With a change of variables to $y\equiv\frac{1-x}{2}$, the integral becomes
    \begin{align*}
        \int_{-1}^{+1}&\dd{x}\qty(\frac{1+x}{1-x})^{\gamma/2}\qty(\frac{x-1}{2})^k\\
        &=2\qty(-1)^k\int_0^{+1}\dd{y}\qty(1-y)^{\gamma/2}y^{k-\frac{\gamma}{2}}.
    \end{align*}
This is a Euler integral of the first kind, and evaluates to
    \begin{equation*}
        \int_0^{+1}\dd{y}\qty(1-y)^{\gamma/2}y^{k-\frac{\gamma}{2}}=\frac{\Gamma\qty(1+k-\frac{\gamma}{2})\Gamma\qty(1+\frac{\gamma}{2})}{\Gamma\qty(k+2)},
    \end{equation*}
    where we assume $0<\gamma<2$.
We therefore have
\begin{widetext}
    \begin{align*}
        C_\ell&=4\uppi\mathcal{A}_\mathrm{gw}\sum_{k=0}^\ell\frac{\qty(-1)^k\qty(\ell+k)!}{\qty(k!)^2\qty(\ell-k)!}\frac{\Gamma\qty(1+k-\frac{\gamma}{2})\Gamma\qty(1+\frac{\gamma}{2})}{\Gamma\qty(k+2)}\\
        &=4\uppi\mathcal{A}_\mathrm{gw}\sum_{k=0}^\ell\qty(-1)^k\frac{\Gamma\qty(\ell+k+1)}{\Gamma\qty(k+1)^2\Gamma\qty(\ell-k+1)}\frac{\Gamma\qty(1+k-\frac{\gamma}{2})\Gamma\qty(1+\frac{\gamma}{2})}{\Gamma\qty(k+2)}\\
        &=4\uppi\mathcal{A}_\mathrm{gw}\sum_{k=0}^\infty\qty(-1)^k\frac{\Gamma\qty(\ell+k+1)}{\Gamma\qty(k+1)^2\Gamma\qty(\ell-k+1)}\frac{\Gamma\qty(1+k-\frac{\gamma}{2})\Gamma\qty(1+\frac{\gamma}{2})}{\Gamma\qty(k+2)},
    \end{align*}
\end{widetext}
    where the final line follows because $\Gamma\qty(\ell+k+1)/\Gamma\qty(\ell-k+1)$ vanishes for $k>\ell$.
We can simplify further by defining the rising Pochammer symbol,
    \begin{equation*}
        \qty(x)_k\equiv x\qty(x+1)\cdots\qty(x+k-1),
    \end{equation*}
    and by using Euler's reflection formula,
    \begin{equation*}
        \mathrm{sinc}\qty(\uppi x)\equiv\frac{\sin\uppi x}{\uppi x}=\frac{1}{\Gamma\qty(1+x)\Gamma\qty(1-x)}.
    \end{equation*}
This gives
    \begin{equation*}
        C_\ell=\frac{4\uppi\mathcal{A}_\mathrm{gw}}{\mathrm{sinc}\qty(\uppi\gamma/2)}\sum_{k=0}^\infty\frac{\qty(-\ell)_k\qty(\ell+1)_k\qty(1-\frac{\gamma}{2})_k}{\qty(1)_k\qty(2)_kk!}.
    \end{equation*}
The series above defines a generalized hypergeometric function ${}_3F_2$, so that we get our final expression,
    \begin{equation}
        C_\ell=4\uppi\mathcal{A}_\mathrm{gw}\frac{{}_3F_2\qty(-\ell,\ell+1,1-\frac{\gamma}{2};1,2;1)}{\mathrm{sinc}\qty(\uppi\gamma/2)}.
    \end{equation}

Note that when $\ell=0$, the hypergeometric function evaluates to unity, and we have
    \begin{equation}
        C_0=\frac{4\uppi\mathcal{A}_\mathrm{gw}}{\mathrm{sinc}\qty(\uppi\gamma/2)}>0.
    \end{equation}
However, the $C_0$ component is just an average of $C_\mathrm{gw}$ over the sphere,
    \begin{align*}
        C_0&=2\uppi\int_{-1}^{+1}\dd{\qty(\cos\theta_\mathrm{o})}C_\mathrm{gw}=\int_0^{2\uppi}\dd{\phi_\mathrm{o}}\int_0^\uppi\dd{\theta_\mathrm{o}}\sin\theta_\mathrm{o}C_\mathrm{gw}\\
        &=\int_{S^2}\dd[2]{\sigma_\mathrm{o}}C_\mathrm{gw},
    \end{align*}
    and therefore must be equal to 0, since the spherical integral and the ensemble averaging process commute, and since the average of $\delta_\mathrm{gw}^{(\mathrm{s})}$ over the sphere is 0 by definition,
    \begin{equation}
        C_0=\ev{\delta_\mathrm{gw}^{(\mathrm{s})}\int_{S^2}\dd[2]{\sigma_\mathrm{o}}\delta_\mathrm{gw}^{(\mathrm{s})}}=0.
    \end{equation}
This is indicative of the inaccuracies inherent to Eq.~\eqref{eq:C-ell-analytical}.

\section{Waveform numerical constants}
\label{sec:waveforms}
The following expressions and numerical values are needed to fully specify the hybrid waveforms given in Eq.~\eqref{eq:waveforms}.
They exactly match those given in Ref.~\cite{Ajith:2009bn}.
\begin{widetext}
\begin{tabular}{@{}llllllll@{}}
    \toprule
    $i\qquad$ & $\nu_i^{0}$ & $y_i^{(10)}$ & $y_i^{(11)}$ & $y_i^{(12)}$ & $y_i^{(20)}$ & $y_i^{(21)}$ & $y_i^{(30)}$\\
    \midrule
    $1$ & $1-4.455\qty(1-\chi)^{0.217}+3.521\qty(1-\chi)^{0.26}$ & $+0.6437$ & $+0.827$ & $-0.2706$ & $-0.05822$ & $-3.935$ & $-7.092$\\
    $2$ & $[1-0.63\qty(1-\chi)^{0.3}]/2$ & $+0.1469$ & $-0.1228$ & $-0.02609$ & $-0.0249$ & $+0.1701$ & $+2.325$\\
    $3$ & $[1-0.63\qty(1-\chi)^{0.3}]\qty(1-\chi)^{0.45}/4$ & $-4.098$ & $-0.03523$ & $+0.1008$ & $+1.829$ & $-0.02017$ & $-2.87$\\
    $4$ & $0.3236+0.04894\chi+0.01346\chi^2$ & $-0.1331$ & $-0.08172$ & $+0.1451$ & $-0.2714$ & $+0.1279$ & $+4.922$\\
    \bottomrule
\end{tabular}

\begin{align*}
    \begin{split}
        \alpha_2&=-\frac{323}{224}+\frac{451}{168}\qty(\frac{\mathcal{M}}{M})^{5/3},\\
        \epsilon_1&=1.4547\chi-1.8897,\\
        c_1&=\nu_1^{-1}\qty[\frac{1+\sum_{i=2}^3\alpha_i\qty(\uppi GM\nu_1)^{i/3}}{1+\sum_{i=1}^2\epsilon_i\qty(\uppi GM\nu_1)^{i/3}}]^2,
    \end{split}\!\!\!\!\!\!\!\!\!\!\!\!\!\!\!\!\!\!\!\!\!\!\!\!\!\!\!\!\!\!\!\!\!\!\!\!
    \begin{split}
        \alpha_3&=\qty[\frac{27}{8}-\frac{11}{6}\qty(\frac{\mathcal{M}}{M})^{5/3}]\chi,\\
        \epsilon_2&=-1.8153\chi+1.6557\\
        c_2&=c_1\nu_2^{-4/3}\qty[1+\sum_{i=1}^2\epsilon_i\qty(\uppi GM\nu_1)^{i/3}]^2.
    \end{split}
\end{align*}
\end{widetext}

\bibliography{astro-sgwb}
\end{document}